\documentclass{astlb}

\usepackage[utf8]{inputenc}
\usepackage[english]{babel}
\usepackage{graphicx}
\graphicspath{ {img/} }
\usepackage{float} 
\usepackage{wrapfig}
\usepackage[dvipsnames]{xcolor}
\usepackage{bm}
\usepackage{amsmath}
\DeclareMathOperator*{\argmax}{argmax}
\usepackage{caption, subcaption}
\usepackage{booktabs}
\usepackage{siunitx}
\usepackage{ulem}
\usepackage{amssymb}
\usepackage{tikz}
\usepackage{mathtools, cuted}
\usepackage{calrsfs}
\usepackage{natbib}
\usepackage{xspace}
\usetikzlibrary{trees}
\usetikzlibrary{arrows.meta, chains, positioning, quotes}
\DeclareMathAlphabet{\pazocal}{OMS}{zplm}{m}{n}
\usepackage[hyperfootnotes=false]{hyperref}

\newcommand{\La}{\mathcal{L}}
\newcommand{\Lb}{\pazocal{L}}

\usepackage{float}
\usepackage{multicol}
\usepackage{url}

\newcommand{\erosita}{{eROSITA}\xspace}
\newcommand{\srge}{{SRG/eROSITA}\xspace}

\newcommand{\xdim}{erg/s/cm$^2$}

\begin{document}

\journalinfo{2021}{0}{0}{1}[0]

\title{SRG\MakeLowercase{z}: building an optical cross-match model for the X-ray \srge sources using the Lockman Hole data}

\author{M. I. Belvedersky\address{1,4}\email{mike.belveder@gmail.com},
A. V. Meshcheryakov\address{1,2}, M. R. Gilfanov\address{1,3}, P. S. Medvedev\address{1}
  \addresstext{1}{Space Research Institute (IKI), Russian Academy of Sciences, Profsoyuznaya ul. 84/32, Moscow, 117997 Russia}
  \addresstext{2}{Lomonosov Moscow State University (MSU), Moscow, Russia}
  \addresstext{3}{Max-Planck-Institut für Astrophysik (MPA), Karl-Schwarzschild-Str. 1, D-85741 Garching, Germany}
  \addresstext{4}{National Research University Higher School of Economics (HSE), Moscow, Russia}
}

\shortauthor{Belvedersky et al.}

\shorttitle{Optical cross-match of the \srge Lockman Hole sources}

\submitted{28.12.2021}

\begin{abstract}

We present a probabilistic  model built for the optical cross-match between the \srge X-ray sources and photometric data from the DESI Legacy Imaging Surveys. The model relies both on positional and photometric information on optical objects nearby X-ray sources and allows performing selection with precision and recall $\approx94$\% (for $F_{\rm X,0.5-2}>10^{-14}$~\xdim{}). With this model, we calibrated  positional error of the \srge sources detected in the Lockman Hole: $\sigma_{\rm corr} = 0.87\sqrt{ \sigma_{\rm det}^{2.53} + 1.12^2}$.

The model will become a part of the SRGz system for data analysis of the X-ray data obtained from the all-sky \srge survey.

\end{abstract}

\setcounter{section}{0}
\setcounter{subsection}{1}

\section{Introduction}

High-energy astrophysics space observatory Spektr-RG (SRG, \citealt{2021A&A...656A.132S}) was launched on July 13, 2019 and is operating  in a halo orbit around the L2 libration point of the Earth-Sun system. The main goal of the observatory is to perform a four-year all-sky survey in the energy band 0.2-30 keV. There are two telescopes onboard SRG: \srge \citep{predehl2021} and Mikhail Pavlinsky ART-XC \citep{pavlinsky2021}.

We can obtain just a few parameters for most of the \srge sources: a position on the sky (within the positional error) and a flux. We need more information to explore physical properties of the sources. First, we need to define their physical class: a star, a QSO, etc. In most cases, we obtain such information from the optical spectrum of the X-ray sources counterparts. To perform optical spectroscopy, one needs to match an X-ray source with its optical counterpart. In other words, to perform the optical cross-match. X-ray sources are usually hard to cross-match because of their significant positional error: often, we see numerous optical candidates within the error radius. This problem gets more complicated within the galaxy plane, where we see numerous optical objects. For these reasons, we need a cross-match model that allows measuring the probability of correct identification. For every optical object in the vicinity of an X-ray source, we need to calculate the probability to be a counterpart $p_{\rm match}$. The model also should allow calculating the probability that an X-ray source does not have an optical counterpart at all within some photometric survey ($p_{\varnothing}$).

The model should respect the following requirements:

\begin{enumerate}
    \item A counterpart prediction must be based on both positional and photometric information on optical objects nearby the X-ray sources
    \item An optical counterpart might be absent in a photometric catalogue. The model must warn us about it with the high value of the according probability ($p_{\varnothing}$).
    \item We are interested in two different selection regimes: with a high precision (80\% and higher) or with a high recall (90\% and higher). The former regime is preferable when we search for unique and rare sources; the latter suits the tasks similar to QSO luminosity function measurement (when one needs to reduce the selection effects). Therefore, the model must allow performing a selection of the X-ray sources in both regimes by varying the $p_{\varnothing}$ threshold.
\end{enumerate}

The most widespread approach to the correlation between astronomical catalogues is based on the likelihood ratio method and was first introduced in \citealt{1992MNRAS.259..413S}. The approach was applied and developed in \citealt{2005AJ....130.2019O, 2013ApJS..209...30N} etc. The likelihood ratio method was successfully applied to find counterparts of the X-ray sources in such surveys as XMM-COSMOS \citep{2007ApJS..172..353B}, Chandra-COSMOS \citep{2012yCat..22010030C, 2019yCat..18170034M}, STRIPE-82X \citep{2016yCat..18170172L, 2019yCat..18500066A}. As a generalisation of the likelihood ratio method, one can consider methods based on Bayesian statistics \citep{2015AnRSA...2..113B, 2017A&A...597A..89P, 2018MNRAS.473.4937S}.

We present a cross-match model built for the correlation of point-like X-ray sources discovered by \srge in the Lockman Hole area. The Lockman Hole (LH, \citealt{1986ApJ...302..432L}) is a sky field that fits well for the observations of extragalactic sources. The absorption of the X-ray radiation is relatively low (the typical column density of neutral hydrogen in this region is about $N_H \approx 4.5 \times 10^{19}$ cm$^{-2}$). With the \srge X-ray data and optical observations from the DESI Legacy Imaging Surveys obtained in this unique area, we have tested a cross-match approach that will be applied for the all-sky \srge survey.

The optical cross-match model for point-like X-ray sources presented in this paper is part of a more complex system called SRGz. The SRGz system has been developed by IKI RAS to analyse the \srge survey. This paper is the first in a series of publications presenting models and algorithms built for the SRGz. The main goal of SRGz is to perform a multi-survey cross-correlation and determine the physical properties of the \srge sources.

One of the main model characteristics is the relationship between the X-ray sources positional error ($\sigma$) and its detection likelihood\footnote{$\Lb=-\ln(p)$, where $p$ is a null hypothesis probability for a source to be produced by the noise component on an X-ray image} ($\Lb$). We will measure the $\sigma(\Lb)$ relationship as a byproduct of the model development using the optical data obtained in an X-ray sources' localisation area.

The standard algorithm for the point-like X-ray sources detection and characterisation allows calculating both position and positional error ($\sigma_{\rm det}$) for every X-ray source. The model we have built can be used for an independent verification of this positional error.

The paper has the following structure. Optical and X-ray data are described in \S\ref{chap:data}. This Section also contains the description of the test sample designed to verify our model. In \S\ref{chap:model} we present our cross-match model built for the correlation of the \srge Lockman Hole X-ray sources. The results are listed in Section \ref{results}. The last Section contains summarised findings and conclusions.

\section{Data} \label{chap:data}

\subsection{X-ray data}

The Lockman Hole observations took place during the performance verification phase of \srge in October 2019. The observations were done in a raster scanning regime. This regime has notable advantages in comparison to the mosaic one, which is the most common scanning regime for the majority of modern X-ray observatories (when survey fields are covered by pointing observations). The raster scanning regime allows obtaining wide field X-ray images with almost unaltered sensitivity (within the image) and PSF.

The raster scanning scheme for the Lockman Hole consists of parallel scans in alternate directions (with the orientation roughly corresponding to the ecliptic coordinate system) with the shift equal 11.7 arcminutes. Scan speed was 9.1 arcseconds (which is an order of magnitude slower than in the whole-sky scanning regime). The field of the view of \erosita is 1 degree which means that every source was observed continuously during $\sim6.6$ minutes. The footprint of the LH survey is $\approx18.5$ square degrees ($5^\circ\times3.7^\circ$) with the centre coordinates $\alpha=10^{\rm h}35^{\rm m}$ and $\delta=+57^{\circ}38^{\prime}$. Total duration of the survey is $180$~ks, mean exposure time is about $8$~ks per point. These parameters allow achieving sensitivity $\approx3\times10^{-15}$~\xdim{} in $0.5 - 2$~keV energy range.

The primary data processing has been performed using the software developed in IKI RAS which includes components of the eROSITA Science Analysis Software System (eSASS, developed in Max Planck Institute for Extraterrestrial Physics, Germany). The source detection was performed using an approximation of a distribution in counts based on the \erosita PSF (the ermldet program, eSASS). The total number of detected sources is $8309$ (with detection likelihood $\Lb>6$). The average surface density in the Lockman Hole is $\sim370$ sources per square degree which is comparable with the density in other X-ray surveys with close characteristics such as \textit{XBootes} (9.3 square degrees, \citealt{2005ApJS..161....1M}) and \textit{XMM-XXL-North} (18.5 square degrees, \citealt{2016yCat..74570110M}).

About 20\% of the survey footprint were observed previously with other X-ray telescopes such as \textit{ROSAT}, \textit{Chandra} and \textit{XMM-Newton}. In this work we use observations obtained by \textit{Chandra} and \textit{XMM-Newton} presented in CSC 2.0 \citep{2010ApJS..189...37E} and 4XMM DR10 \citep{2020A&A...641A.136W} catalogues accordingly. These data allow us to examine the quality of our cross-match algorithm (see \S\ref{chap:test} bellow).

\subsection{Optical data}

To correlate the X-ray \srge sources, we will use optical data from public photometric surveys --- the DESI Legacy Imaging Surveys DR8 (DESI LIS, \citealt{2019AJ....157..168D}) and SDSS DR14 \citep{2018ApJS..235...42A}.

\begin{figure}
    \centering
    \includegraphics[width=\linewidth]{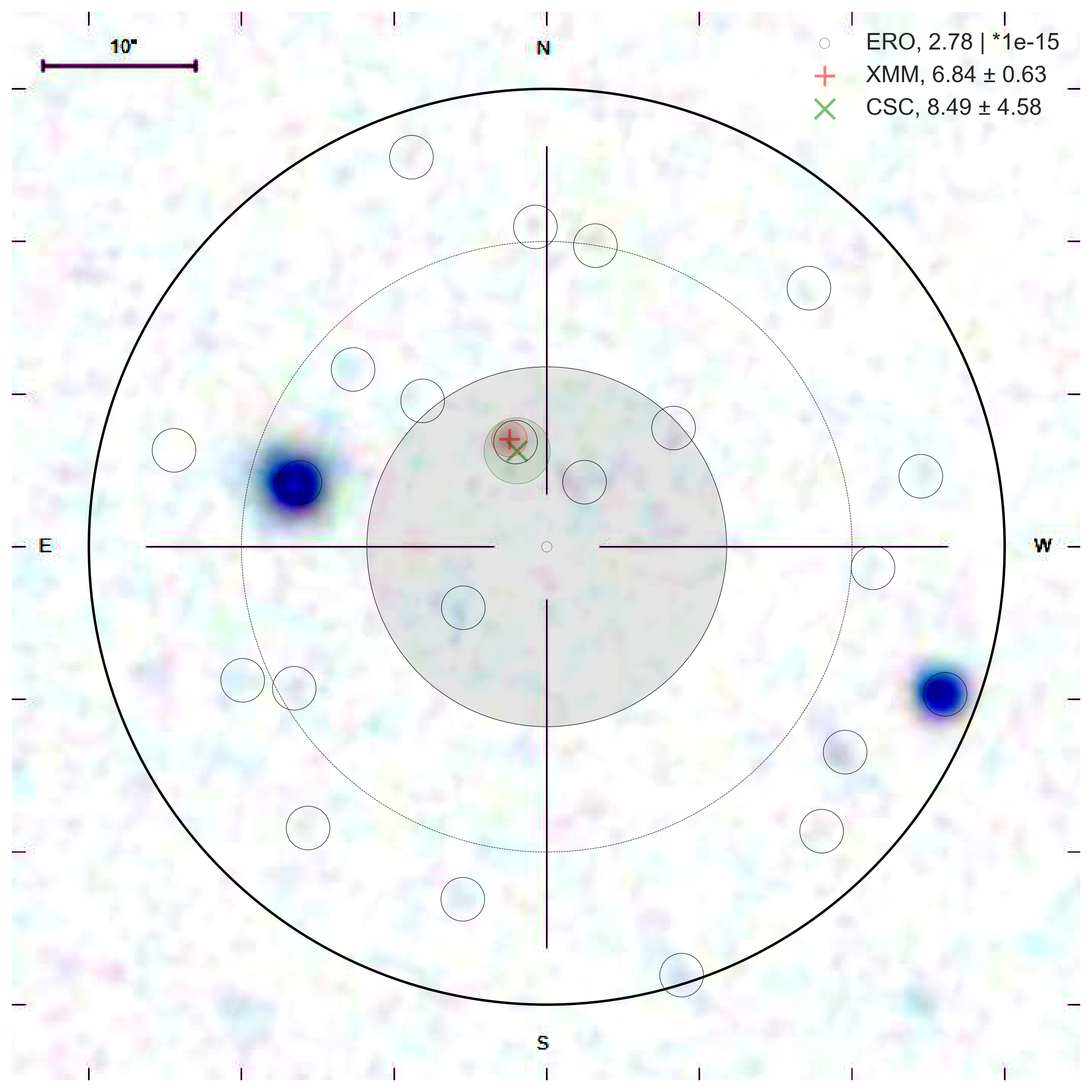}
    \begin{center}
    \caption{An example of an optical field around the \srge source from the test sample with counterparts. Semitransparent circles are the 95\% localisation areas of the X-ray sources \srge (a grey region in the centre), XMM-Newton (an area around the "+" symbol), and Chandra (an area around the cross). The XMM-Newton and Chandra data indicate an optical counterpart (see details in the text).
    \label{fig:dmag_scheme}}
   \end{center}
\end{figure}

\subsection{Test catalogue with optical counterparts for the Lockman Hole} \label{chap:test}

To examine the quality of our model, we prepared the test sample of the X-ray sources previously observed by the Chandra and XMM-Newton telescopes (from now on, test sample with optical counterparts). We gathered several hundred \srge sources that can be reliably correlated with their optical counterparts using sources from the CSC 2.0 (Chandra Source Catalog Release 2.0) and 4XMM DR10 (XMM-Newton X-ray source catalogue, data release 10). To simulate optical fields around the \srge sources with no optical counterparts in the DESI LIS, we used the test sample with counterparts after removing all optical counterparts from it (test sample without counterparts).

Test sample creation procedure will be described below.

\begin{enumerate}

    \item In total, \srge detected 8309 point-like sources. First, we choose those that have either one CSC source (585 cases) or one 4XMM source (788) within \ang{;;30}. From these CSC and XMM sources we choose those that have only one DESI LIS object nearby (within the search radius $r~<~r_{\rm false}$\footnote{The search radius \(r_\text{false}\) was calculated as \(r_\text{false}~=~(-\ln[1-f_\text{thresh}]\pi^{-1}\rho_\text{desi}^{-1})^{1/2}~=\ang{;;1.43}\), where \(\rho_\text{desi}~\approx~4.7\times 10^{-3}~\text{arcsec}^{-2}\) is an average sky density of the DESI LIS objects within LH, \(f_\text{thresh}=0.03\) is the probability to find one source or more by chance within \(r<r_\text{false}\) according to the Poisson distribution with \(\lambda=\rho_\text{desi}\)}). Thus, we guarantee that the test sample purity is at least 97\% (it is higher in practice due to the additional selection criteria). At this point the amount of the selected sources is 437 CSC and 473 4XMM sources.
    
    \item Then we left CSC and 4XMM sources with an X-ray flux that differs by the factor of no more than five from the flux of the \srge sources closest to them. At this point, 783 sources were selected in total (319 and 464 optical counterparts based on the CSC and 4XMM data accordingly).
    
    \item Further, we excluded fields containing several 4XMM sources besides the single CSC source and vice versa (cutting the total number of counterparts in the sample to 577). Moreover, after a visual inspection, we decided to exclude the \srge sources with bright (magnitude in $r$ or $z$ filter less than 16) DESI LIS objects within \ang{;;40} around them. We did this to clean the test sample from the DESI LIS optical fields with bright star boguses on the optical image. As a result, the final test sample contains 541 X-ray sources with reliable optical counterparts.
    
\end{enumerate}

Fig.~\ref{fig:dmag_scheme} shows an example of the optical field around the \srge source from the test sample with counterparts. Numerous DESI LIS objects (small circles) are inside the bigger circle with a radius $R_{\rm match}=\ang{;;30}$ (the bold solid line) with an X-ray source in its centre. An optical image in the background was taken\footnote{We used SciServer API for Python \url{https://github.com/sciserver/SciScript-Python}} from the SDSS DR16. \srge source 95\ localisation area is shown by the solid circle in the centre. Similar areas for the 4XMM DR10 and CSC 2.0 sources are also present (marked by the plus and cross signs, respectively). Fluxes for all the mentioned X-ray sources are shown in the legend (in \xdim{}). Both the XMM-Newton and Chandra observations point at the \srge source's optical counterpart.

\begin{figure}
    \centering
    \includegraphics[width=\linewidth]{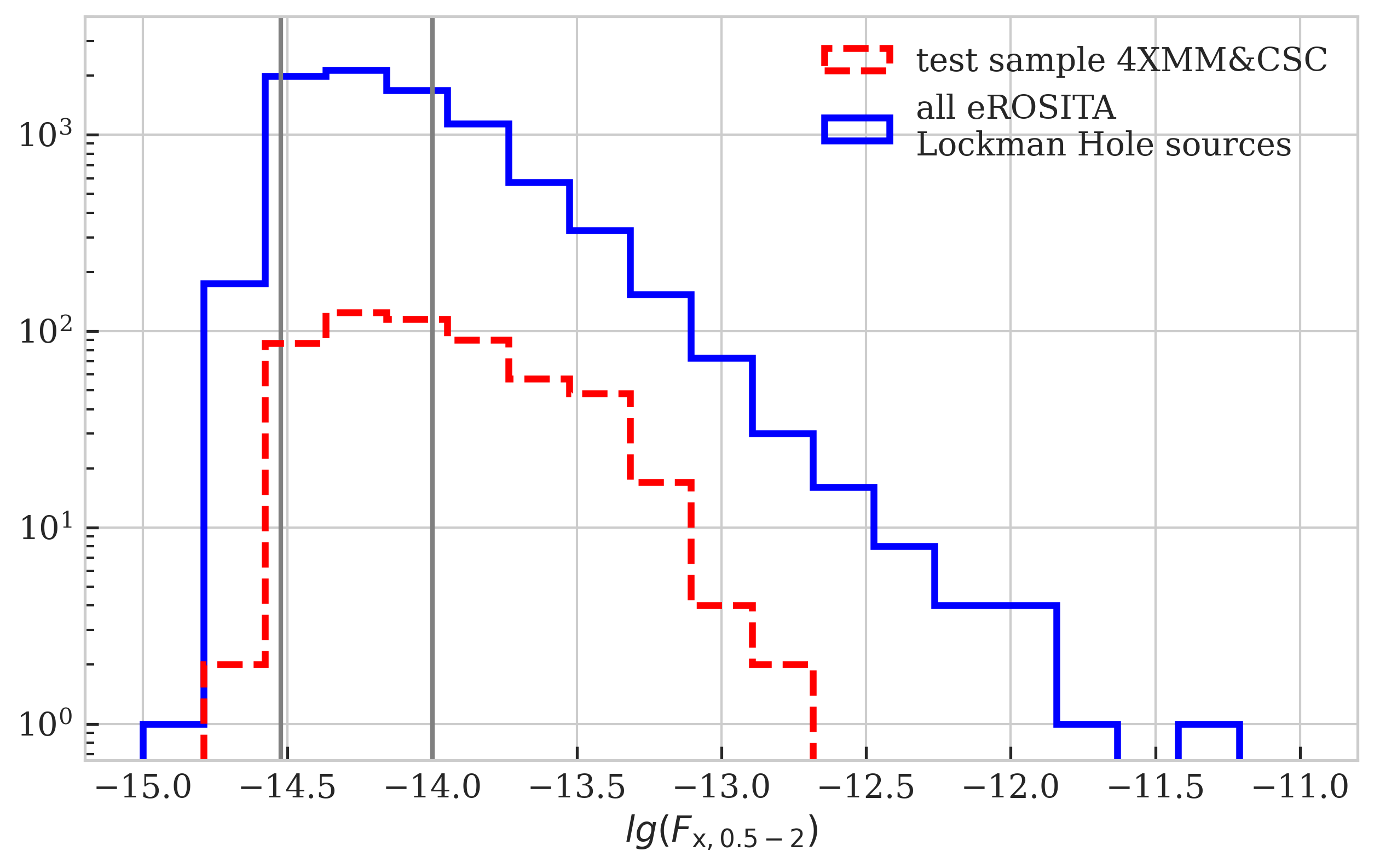}
    \begin{center}
    \caption{An X-ray flux distribution for all the \srge Lockman Hole sources (the solid line), a flux distribution for the test sample X-ray sources (dashed line). Vertical lines mark sensitivity threshold values \num{3e-15} and $10^{-14}$ \xdim{} (LH threshold sensitivity and a four-year sensitivity accordingly)}
    \label{fig:pseudo_reason}
    \end{center}
\end{figure}

Fig.~\ref{fig:pseudo_reason} shows the X-ray flux ($0.5-2$\, keV) distribution for all the \srge Lockman Hole sources and the flux distribution for the test sample X-ray sources. One can see that fluxes from these samples are distributed slightly differently.

Finally, the Lockman Hole test catalogue of reliable optical counterparts is based on the test sample with counterparts. The test catalogue contains all the \srge LH sources paired with the optical fields taken from the test sample with counterparts. We used a bootstrap procedure to get an optical field for an X-ray source. We also used the dependency shown in Fig.~\ref{fig:pc_flux_result} by the solid line to predict that an X-ray source has an optical counterpart (in a predefined photometric catalogue). If the simulation showed that an X-ray source $X$ has an optical counterpart, then an optical field was chosen by chance from the test sample with counterparts (but only from the sources with the flux close enough to the $X$'s flux). Otherwise, an X-ray source does not have a counterpart. Then, we paired this source with an optical field sampled from the whole test sample without counterparts (regardless of X-ray fluxes). All the sampling was done with replacement.

\section{Model} \label{chap:model}

\begin{figure*}
    \centering
    \includegraphics[width=0.8\linewidth]{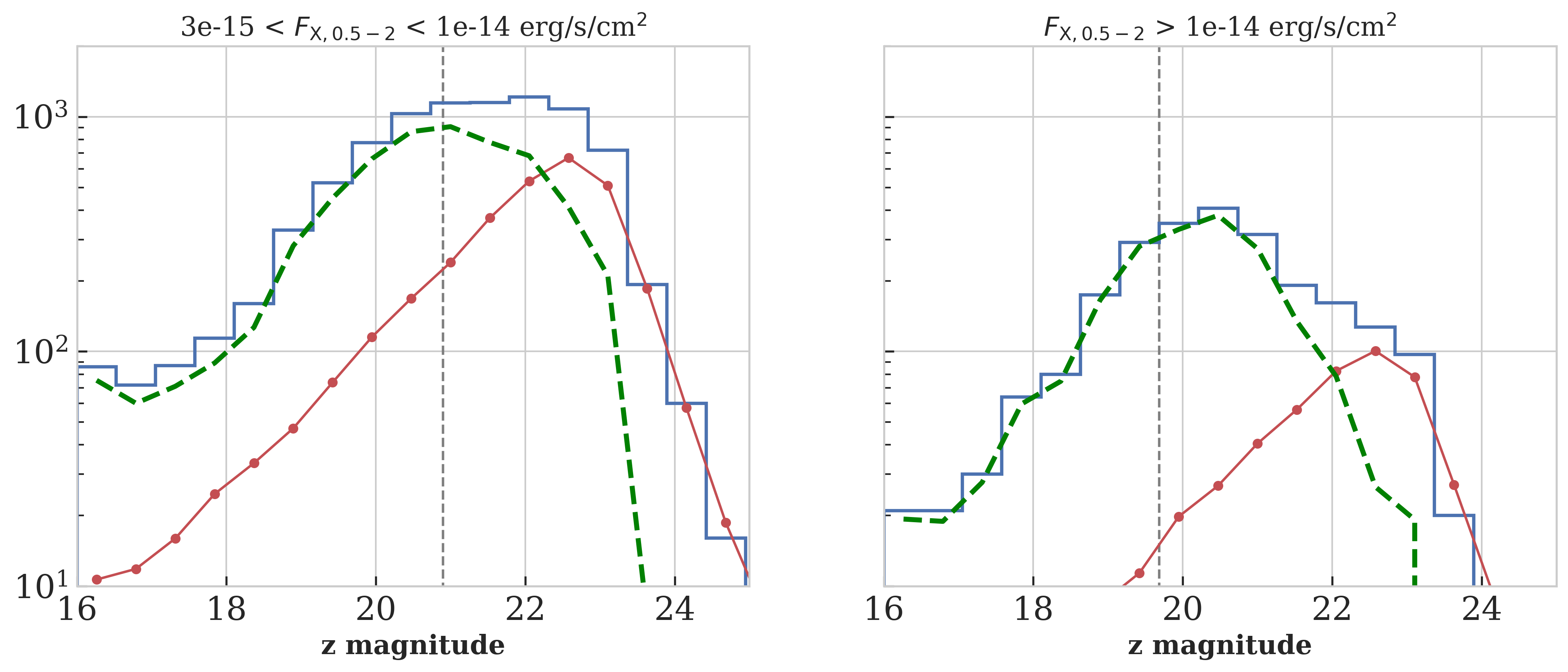}
    \includegraphics[width=0.8\linewidth]{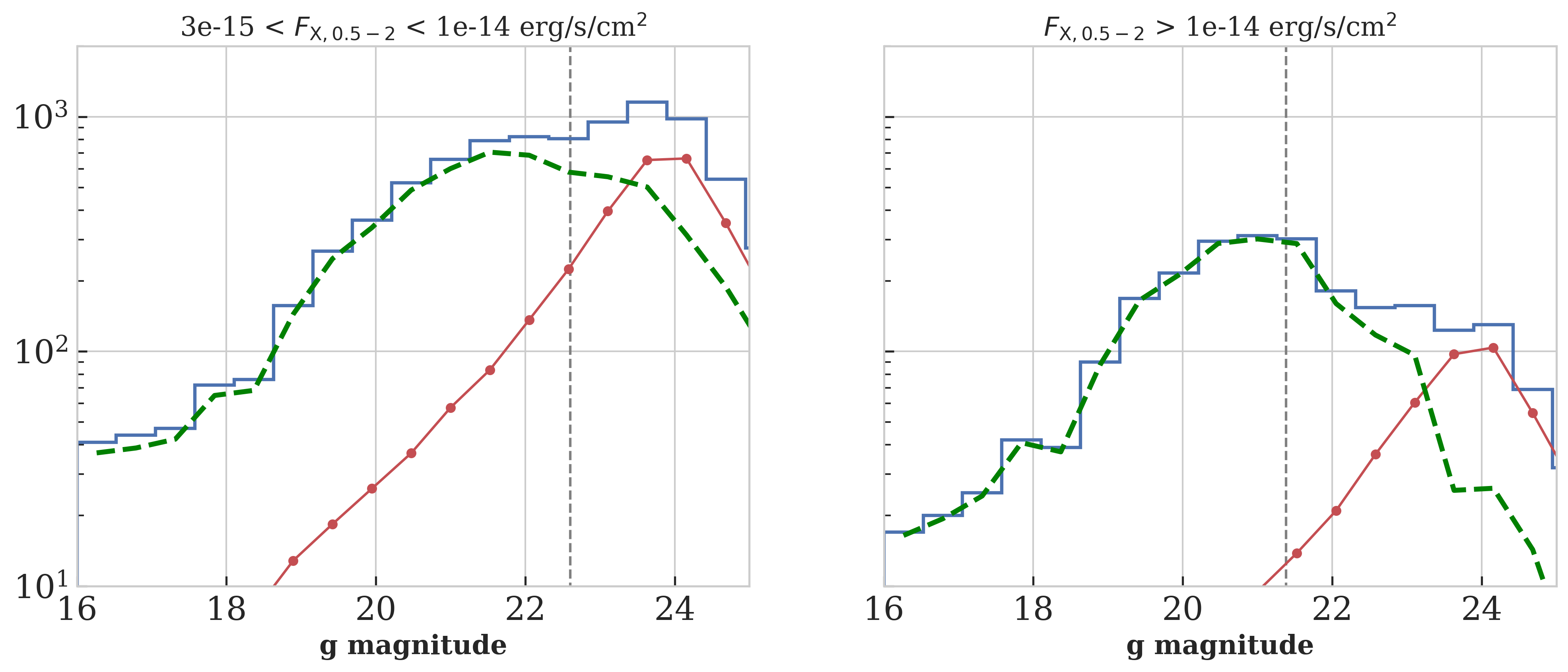}
    \includegraphics[width=0.8\linewidth]{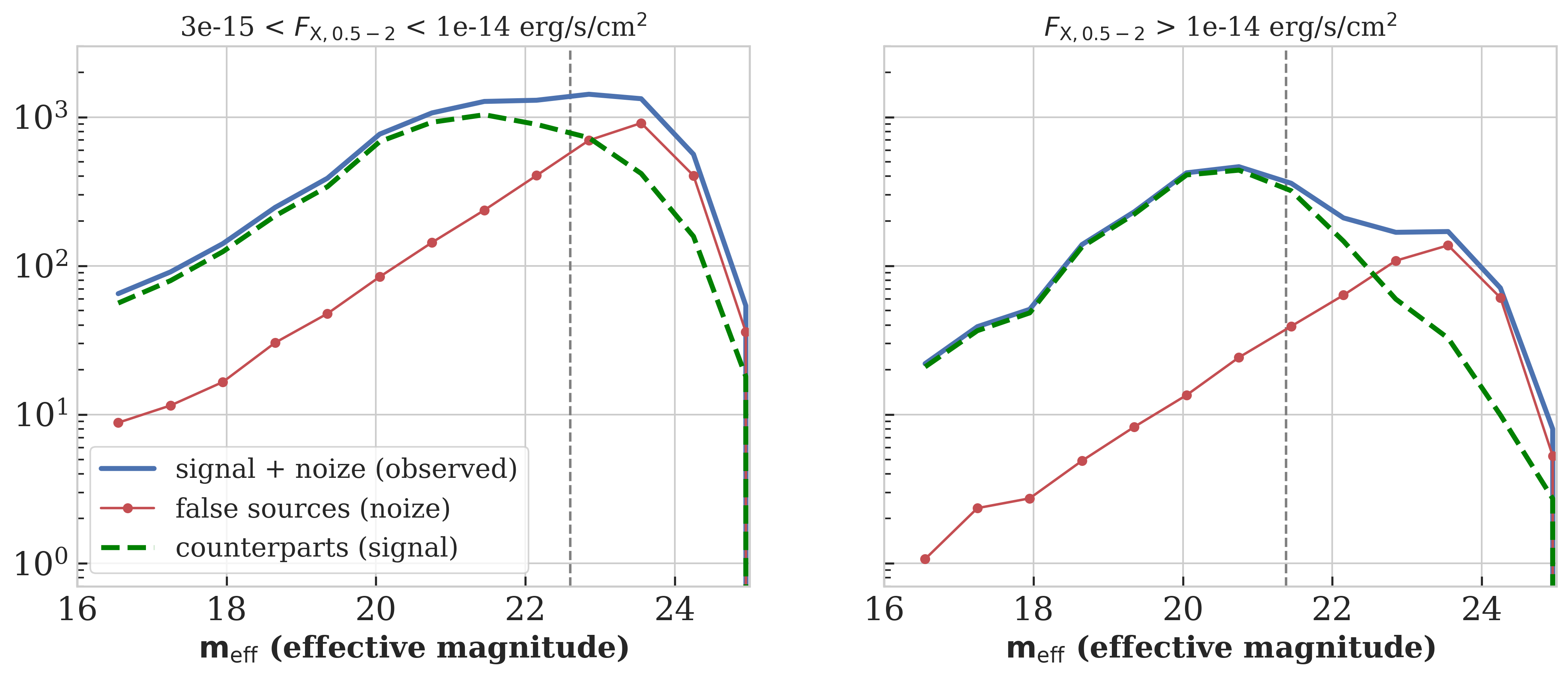}
    \begin{center}
    \caption{The distribution of all optical objects found within the radius $2\sigma$ from the LH sources with the flux $\num{3e-15} < F_{\rm X, 0.5 - 2} < 10^{-14}$ (left graph column) and $F_{\rm X, 0.5 - 2} > 10^{-14}$~\xdim{} (right graph column; see text for more details)}
    \label{fig:dmag_thresh}
    \end{center}
\end{figure*}

For every \srge source, our optical cross-match model has to calculate  1) a probability that this X-ray source has a counterpart in a predefined photometric survey and 2) a probability for every optical candidate to be a counterpart. False optical candidates (objects appearing nearby the X-ray source by chance) and counterparts belong to different distributions in separation (see Fig.~\ref{fig:dmag_thresh}) and in magnitude (see Fig.~\ref{fig:dmag_thresh}, \S\ref{chap:plim_model}). Our model uses photometric information as the \textit{effective magnitude} for the optical objects (see \S\ref{chap:plim_model}).

For each X-ray source ($X$), we calculate the probability $p_{\varnothing}$ that the sample of optical candidates in the $X$'s localisation area \textit{does not} contain an optical counterpart for $X$. For every optical object $o_{\rm i}$, we will find the probability $p_{\rm match,i}$ that it is an optical counterpart for a given X-ray source (taking into account both angular separation and photometric information).

\begin{figure}[H]
    \centering
    \begin{tikzpicture}[level distance=1.5cm,
        level 1/.style={sibling distance=5cm}]
        \node {$X, \{o_{\rm i}\}$}
          child {node {$p_{\varnothing} \geq p'_{\varnothing}$}
            child {node {No counterpart}}
          }
          child {node {$p_{\varnothing} < p'_{\varnothing}$}
            child {node {$\argmax\limits_{o_{\rm i}} p_{\rm match}(o_{\rm i})$}}
          };
    \end{tikzpicture}
    \caption{Counterpart search tree. On level 1 (depending on the threshold value $p'_{\varnothing}$ of the parameter $p_{\varnothing}$) we decide if there is a counterpart among optical candidates. If a counterpart is present, on level 2 for every optical candidate we calculate the probability $p_{\rm match}$ to be a counterpart. A candidate with a largest $p_{\rm match}$ is considered to be a counterpart.}
    \label{fig:cross_scheme}
\end{figure}
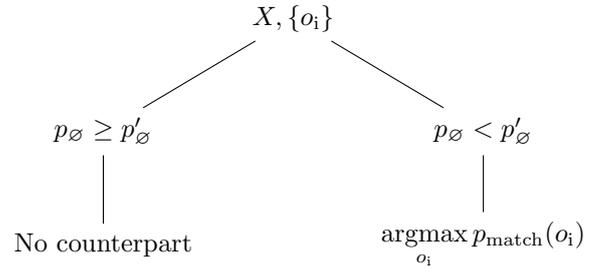

Fig.~\ref{fig:cross_scheme} shows the search tree for the X-ray source optical counterpart. Our algorithm consists of two levels. On the first level, we calculate $p_{\varnothing}$ based on the characteristics of the X-ray source and its optical surroundings. If there has to be a counterpart (if $p_{\varnothing}<p^\prime_{\varnothing}$), then we calculate the probability $p_{\rm match}$ for every optical candidate on the second level. Then we choose the most probable candidate as a counterpart.

In Subsection \S\ref{sec:meff} we introduce effective magnitude for optical objects and illustrate how it helps aggregate photometric information from different filters. In \S\ref{chap:pos_model}, we present our cross-match model. In \S\ref{chap:pmatch_p0}, two expressions are introduced: for the probability $p_{\rm match}$ that a given candidate is a counterpart and the probability $p_{\varnothing}$ that a given X-ray source does not have a counterpart in a predefined photometric survey.

\subsection{Effective magnitudes}\label{sec:meff}

As mentioned above, false optical candidates and counterparts belong to different distributions in separation and magnitude (due to the known correlation between optical and X-ray QSO flux).

To illustrate the concept we plotted the distribution of all optical objects found within the radius $2\sigma$ from the LH sources in a different flux bins: $\num{3e-15} < F_{\rm X, 0.5 - 2} < 10^{-14}$ (left plot panel) and $F_{\rm X, 0.5 - 2} > 10^{-14}$~\xdim{} (right graph column). The solid line is the distribution of all optical objects, the line with circle markers is for false object contributions, the dashed line is the optical counterpart contributions. We measured the number of the false objects between two circles \ang{;;20}--\ang{;;30} (with an X-ray source in their centres). Counterparts' contribution (dashed line) was calculated by extracting the number of false objects from the total number of candidates (taking into account both $2\sigma$-circle and areas between \ang{;;20}--\ang{;;30}). It is clear that as the X-ray flux increases, these two contributions (of false objects and counterparts) to the total distribution become more and more separable. Fig.~\ref{fig:dmag_thresh}, the upper panel shows the magnitude distributions in the $z$ filter, the middle panel is for the $g$ filter. The lower panel shows an effective magnitude distribution (see below). 85\% quantile of the counterparts' distribution by the effective magnitude is shown by the vertical line on the lower panel ($m_{\rm eff} = 22.6$). The vertical lines on the other panels correspond to the values calculated with the expression (\ref{eq:meff}) (where $m_{\rm eff} = 22.6$).

\begin{figure}
    \centering
    \includegraphics[width=\linewidth]{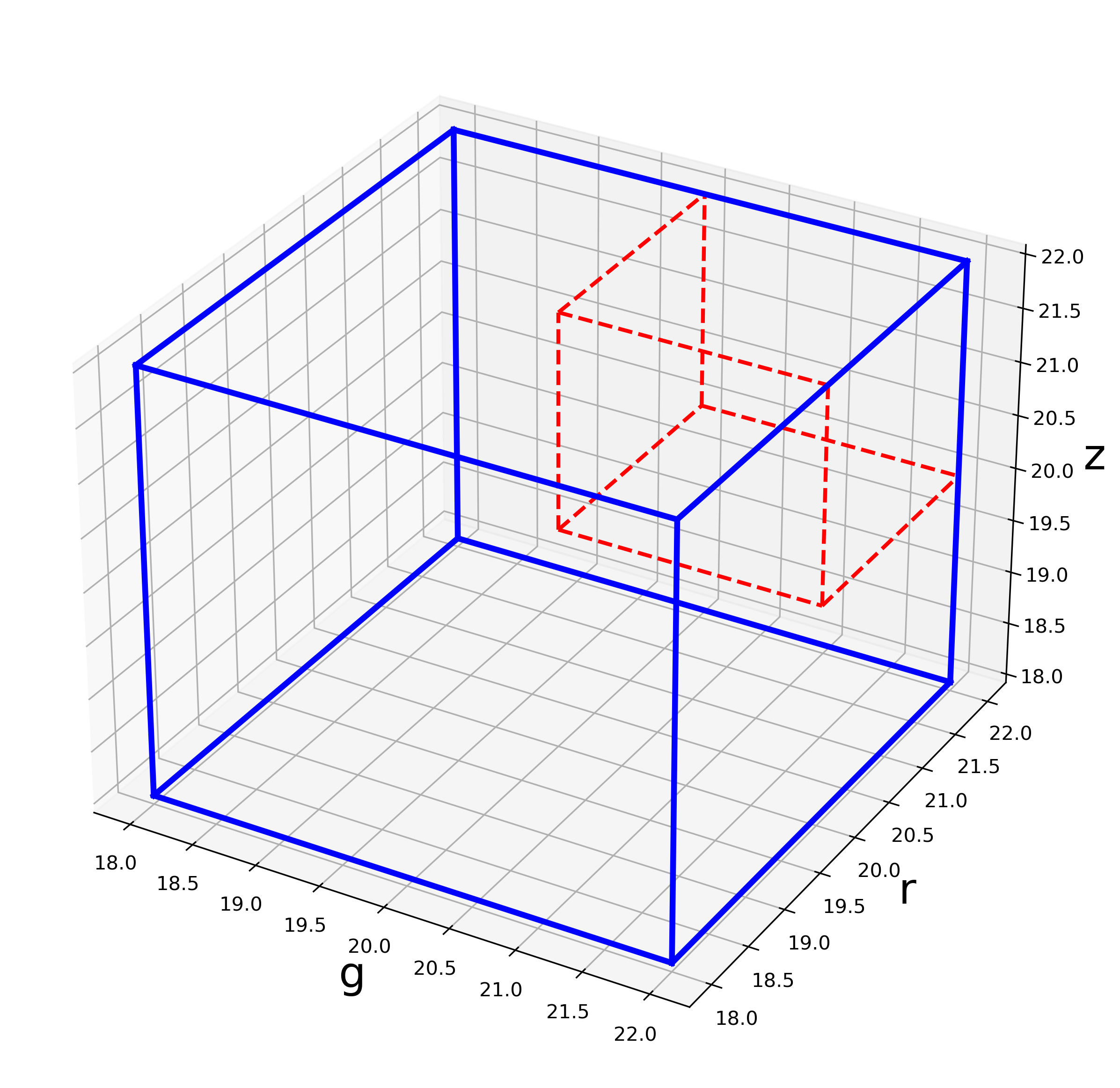}
    \begin{center}
    \caption{Effective magnitude ($m_{\rm eff}$) graphical representation. Magnitudes in the filters $g,\ r$ and $z$ are marked on the cube's edges (the solid lines). The objects with a given $m_{\rm eff}$ are lying on the small cube's faces (bounded by the dashed lines) and their extensions to the bigger photometric magnitudes. The excluded area lies within the dashed cube (and its extension to the bigger photometric magnitudes).}
    \label{dmag_cube}
    \end{center}
\end{figure}

To take into account the photometric information on the optical candidates, we introduce the effective magnitude $m_{\rm eff}$. This parameter is defined by the three optical filters used in the DESI LIS:
\begin{equation}
    m_{\rm eff} = (m_g + c_g) ~|~ (m_r + c_r) ~|~ (m_z + c_z),
    \label{eq:meff}
\end{equation}
where $c_g$, $c_r$, $c_z$ are predefined constants. $m_g$, $m_r$, and $m_z$ magnitudes are AB magnitudes. A graphical interpretation of $m_{\rm eff}$ is shown in Fig.~\ref{dmag_cube}. If an optical candidate has the effective magnitude $m_{\rm eff}$, then its AB magnitude in the $g$ filter is $(m_g + c_g)$ \textit{or} its AB magnitude in the $r$ filter is $(m_r + c_r)$ \textit{or} its AB magnitude in the $z$ filter is $(m_z + c_z)$. The idea behind the effective magnitude is to define the spectral flux density level which an optical object can reach in any of the given filters. In the photometric surveys (such as SDSS, DESI LIS, Pan-STARRS) threshold sensitivity in the red optical filters ($z$) is significantly higher than in the blue ones ($g$). That is why it makes sense to define $c_g$, $c_r$, $c_z$ so that they correspond to the similar signal/noise ratio in the different optical filters. We chose $c_g = 0$, $c_r = 0$, $c_z = -1.7$ which corresponds to threshold sensitivity difference for the photometric data SDSS (\citealt{2018ApJS..235...42A}). Generally, one can consider $c_g$, $c_r$, $c_z$ as the hyperparameters (as it is done in the machine learning applications) and calibrate their values to achieve the optimal result in accordance with a cross-match task.

It is worth mentioning that the effective magnitude allows one to squeeze all photometric information into a single parameter $m_{\rm eff}$. On the one hand, this approach makes photometric information processing a bit imprecise, but on the other hand, it is less complicated and more interpretable. It also makes our model more robust to the noise and systematic errors in the data. Further, we plan to consider more complex approaches based on the machine learning algorithms to perform more flexible and precise photometric information processing.

\begin{figure*}
\centering 
\medskip
\begin{subfigure}{0.3\linewidth}
  \includegraphics[scale=0.25]{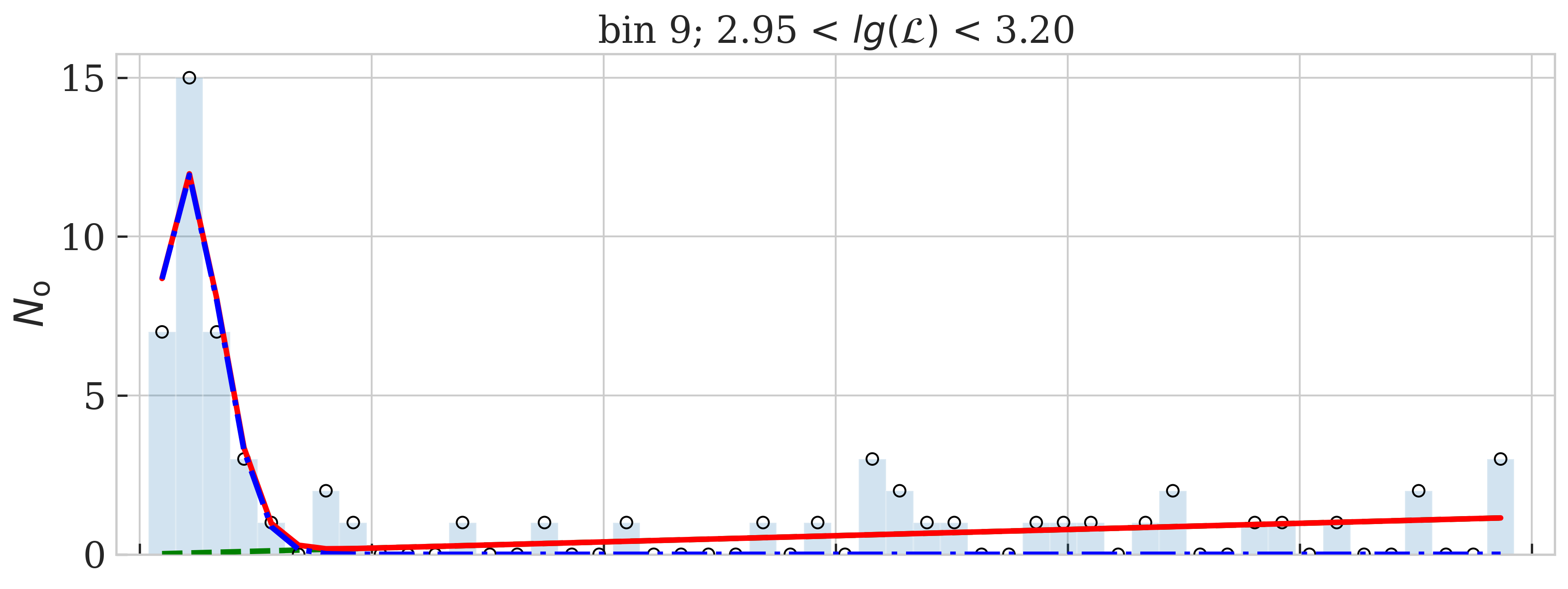}
\end{subfigure}\hfil 
\begin{subfigure}{0.3\linewidth}
  \includegraphics[scale=0.25]{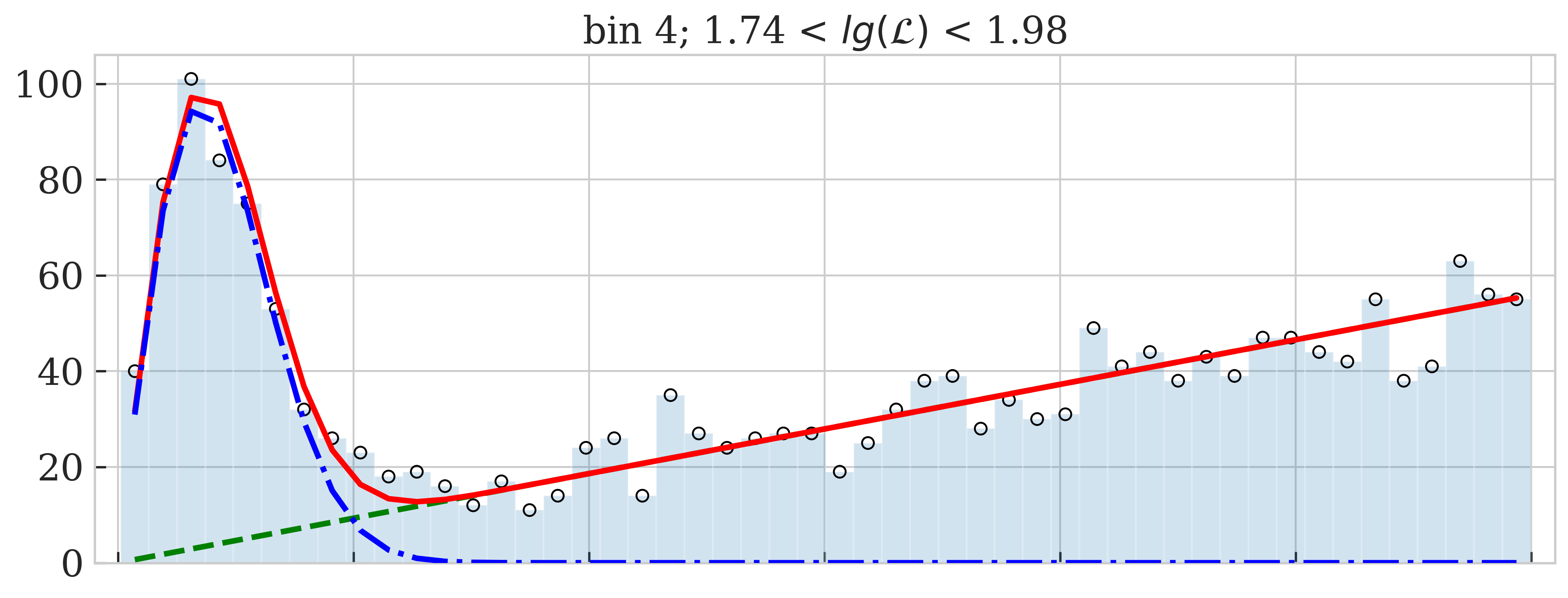}
\end{subfigure}\hfil 
\medskip\\
\begin{subfigure}{0.3\linewidth}
  \includegraphics[scale=0.25]{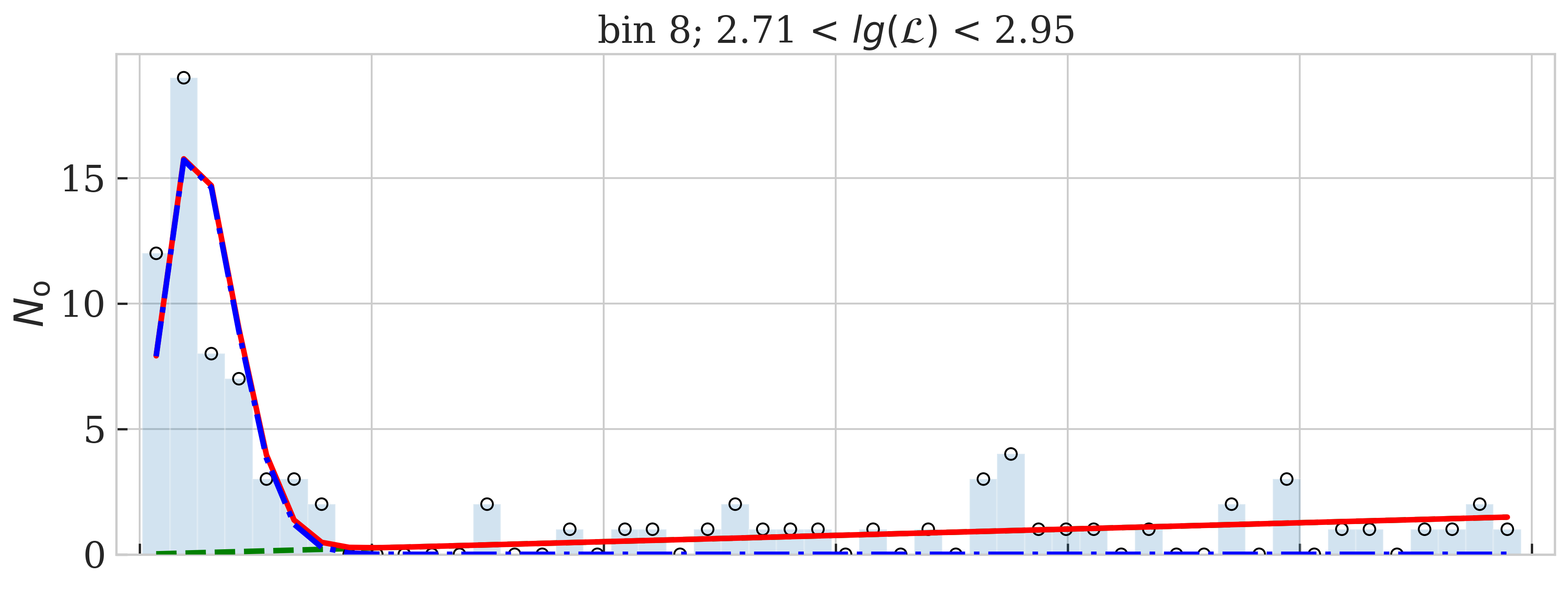}
\end{subfigure}\hfil 
\begin{subfigure}{0.3\linewidth}
  \includegraphics[scale=0.25]{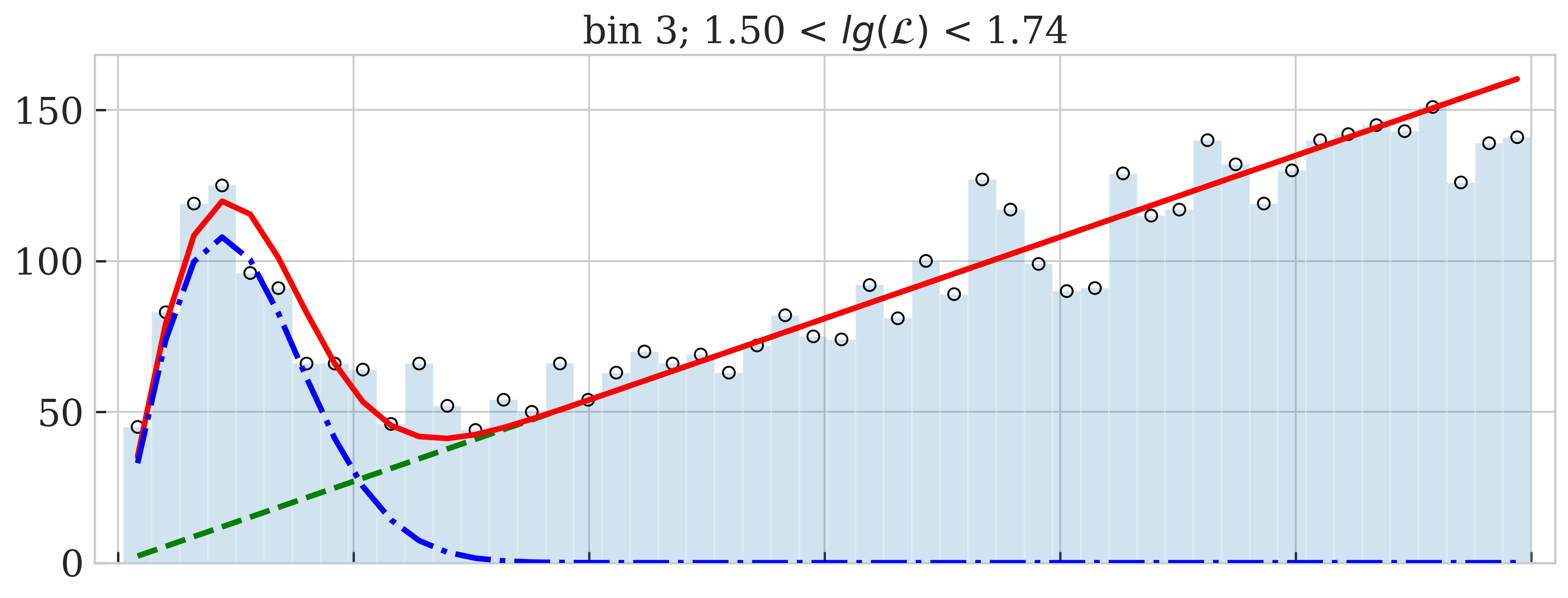}
\end{subfigure}\hfil 
\medskip\\
\begin{subfigure}{0.3\linewidth}
  \includegraphics[scale=0.25]{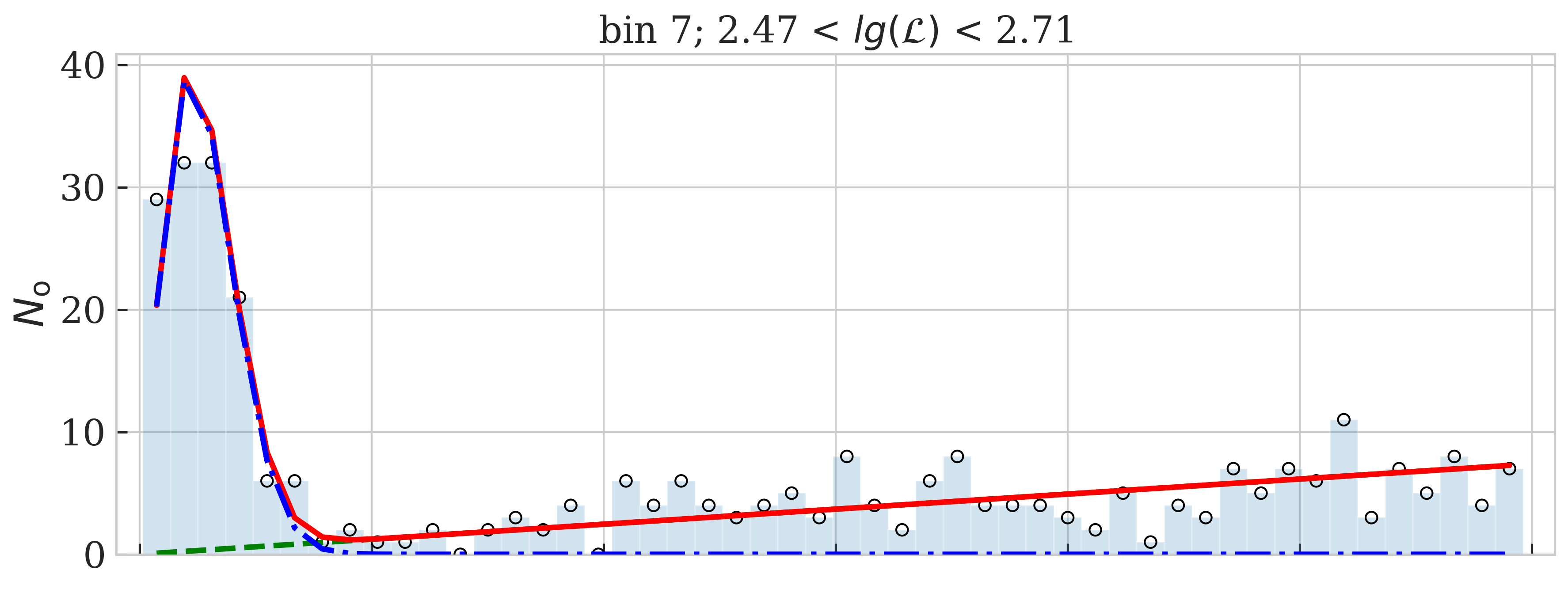}
\end{subfigure}\hfil 
\begin{subfigure}{0.3\linewidth}
  \includegraphics[scale=0.25]{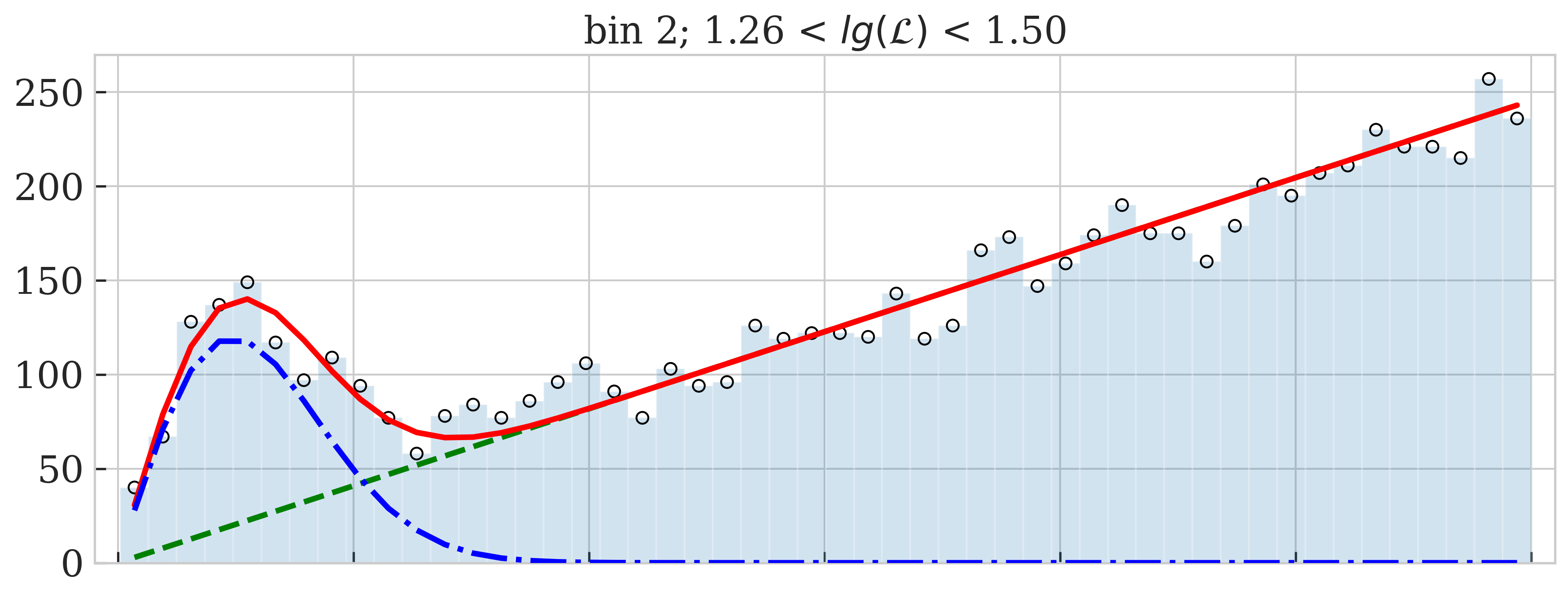}
\end{subfigure}\hfil 
\medskip\\
\begin{subfigure}{0.3\linewidth}
  \includegraphics[scale=0.25]{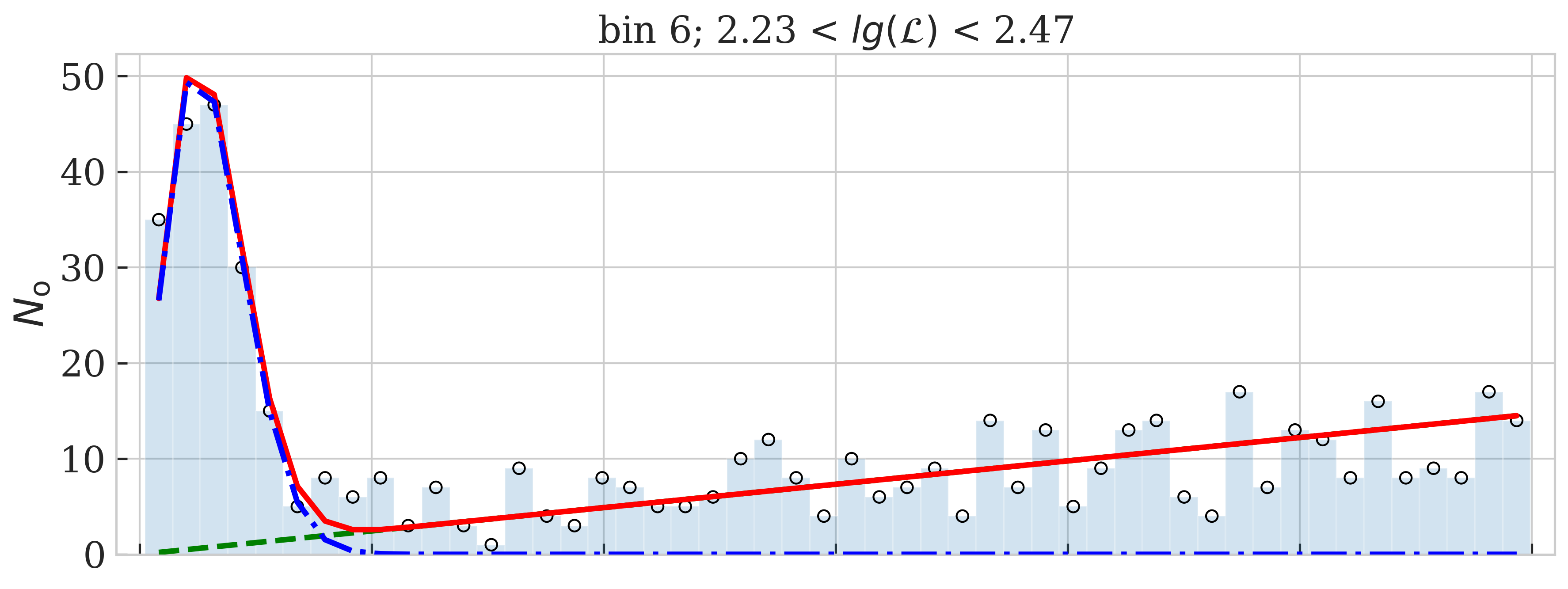}
\end{subfigure}\hfil 
\begin{subfigure}{0.3\linewidth}
  \includegraphics[scale=0.25]{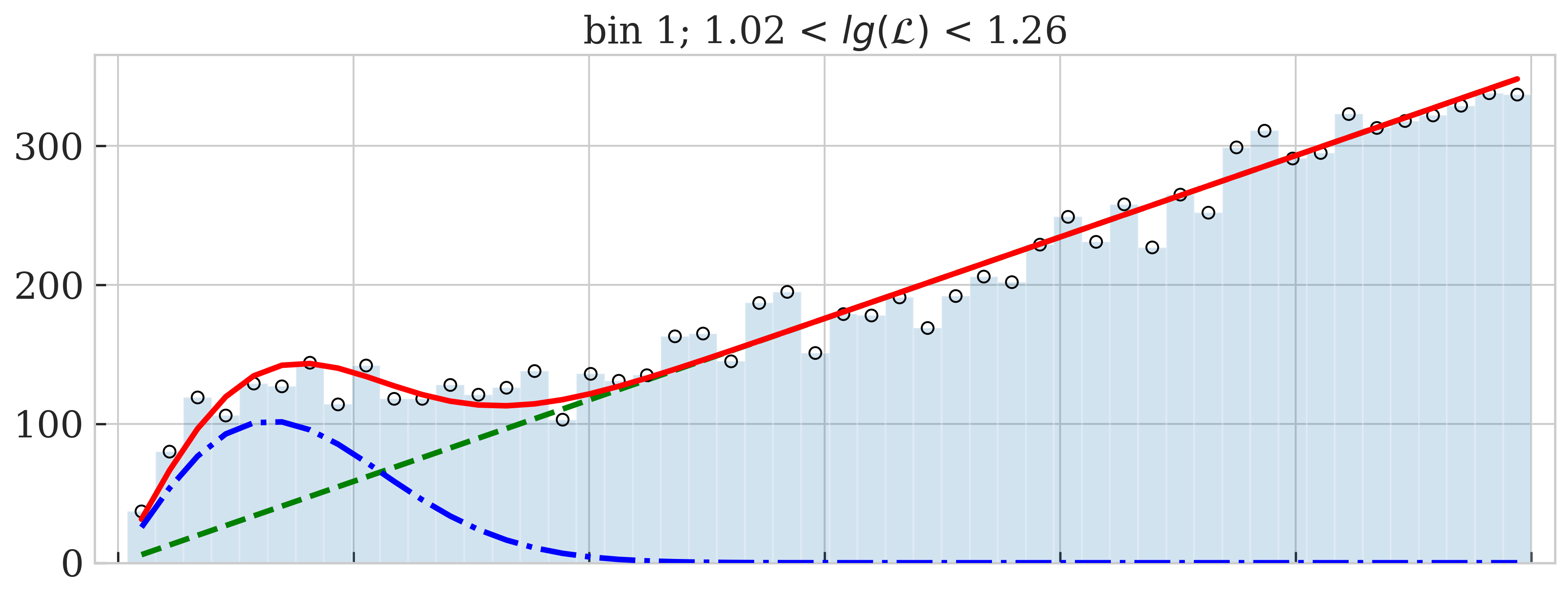}
\end{subfigure}\hfil 
\medskip\\
\begin{subfigure}{0.3\linewidth}
  \includegraphics[scale=0.25]{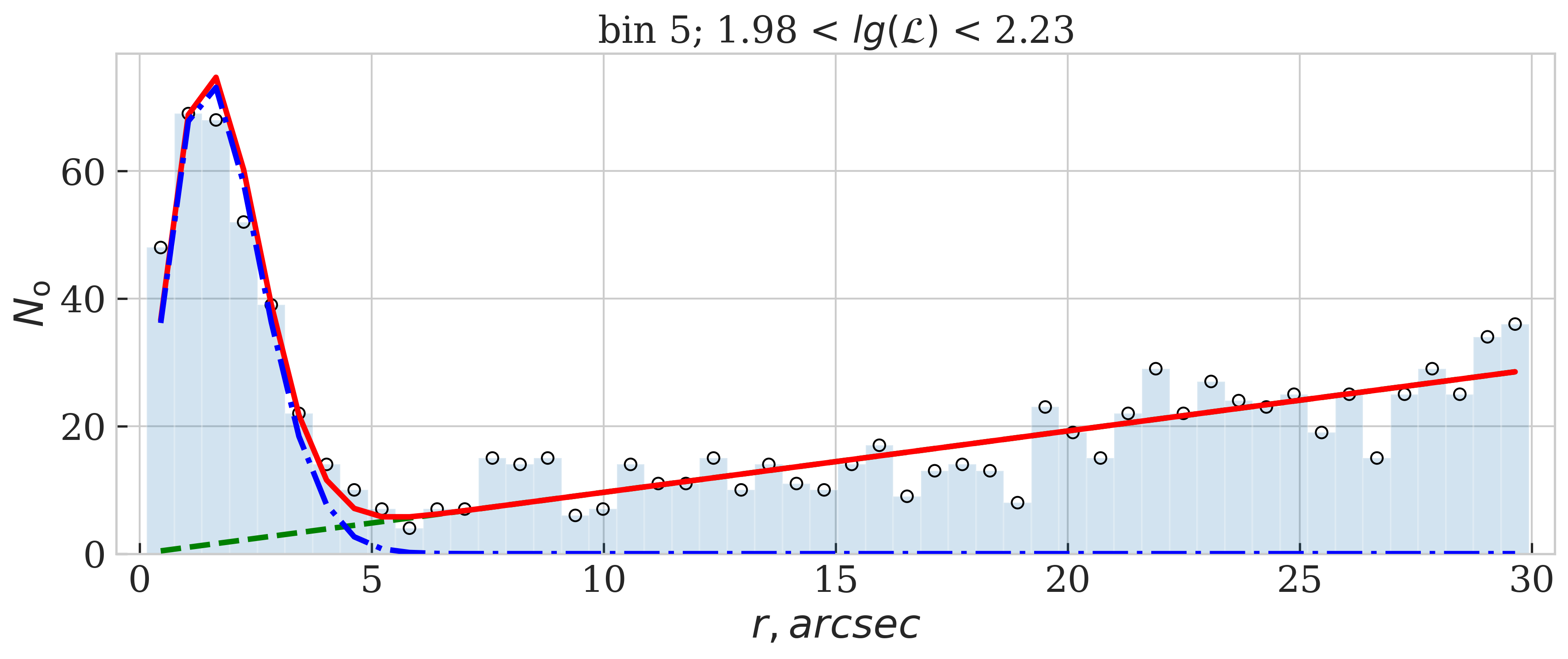}
\end{subfigure}\hfil 
\begin{subfigure}{0.3\linewidth}
  \includegraphics[scale=0.25]{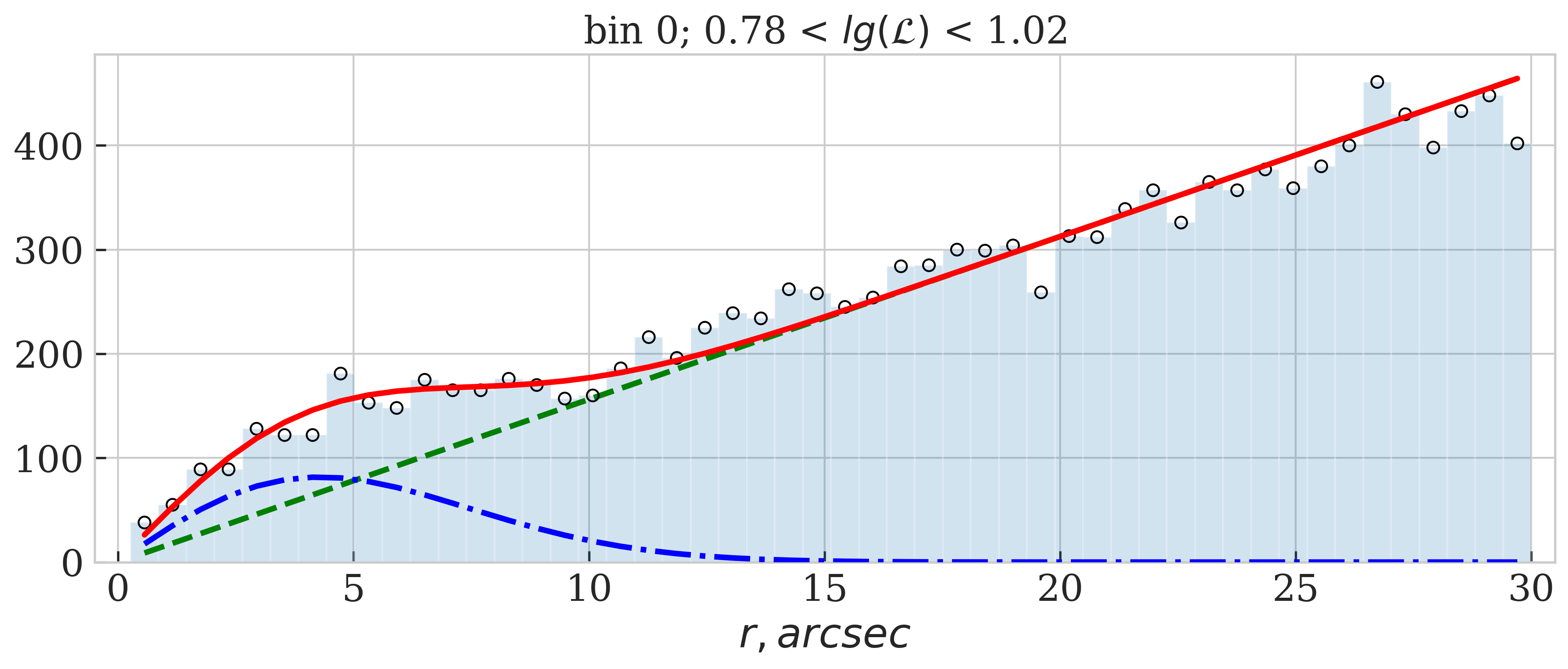}
\end{subfigure}\hfil 
\medskip\\
\caption{The distributions in separation to the optical candidates (empty circles) for $p_{\rm lim} = 0.85$, see \S\ref{chap:plim_model}. The dash-dotted line is the expected distribution of the candidates, the dashed line is the expected distribution of the false objects. The solid line shows the expected total distribution.}
\label{fig:fit_in_bin}
\end{figure*}

\subsection{Positional model} \label{chap:pos_model}

An optical candidate belongs to either candidates distribution or false objects distribution.

We introduce the optical cross-match model with the following constraints (typical in this context; see, for example, \citealt{2015AnRSA...2..113B}):

\begin{itemize}

\item The positional error function of the X-ray sources is described by the two-dimensional normal distribution with the covariance matrix
$\begin{pmatrix}
  \sigma^2 & 0\\ 
  0 & \sigma^2
\end{pmatrix}$.

\item Then the separation $r$ between the true position of an X-ray source and its detected position (for $r\ll\pi$) is described by the Rayleigh distribution:

\begin{equation}
    \phi(r) = \frac{r}{\sigma^2}\exp\bigg(-\frac{r^2}{2\sigma^2}\bigg) ~.
\end{equation}

\item The localisation error of the optical objects (DESI LIS) is negligibly small compared to the X-ray sources localisation error.

\item The false (independent from the X-ray sources) objects distribution is Poisson within the small area in the Lockman Hole (with the surface density $\rho$).

\end{itemize}

Suppose an X-ray catalogue contains $N_{\rm X}$ sources. Then an average number $N_{\rm o}(r)$ of all optical candidates in a circle with the radius $r$ is described by the following expression:

\begin{equation}
    \begin{aligned}[t]
        N_{\rm o}(r) &= N_{\rm x}\Bigg(p_{\rm c} \int_{0}^{r}\phi(r)2\pi r dr + \rho \pi r^2\Bigg) =\\
             &= N_{\rm x}\Bigg(p_{\rm c} \bigg[1 - \exp\bigg(\frac{-r^2}{2\sigma^2}\bigg)\bigg] + \rho \pi r^2\Bigg) ~,
    \end{aligned}
	\label{eq:statistic}
\end{equation}
where $p_{\rm c}(F_{\rm X, 0.5 - 2})$ is the probability for an X-ray source with a given flux to have a counterpart. It is important to underline the difference between $p_{\rm c}$ and $p_{\varnothing}$. $p_{\varnothing}$ is the probability that there is no counterpart for an X-ray source in a given photometric catalogue \textit{considering the vicinity of this X-ray source} (i.e. a specific set of optical objects near this X-ray source); see more details in \S\ref{chap:pmatch_p0}. At the same time, $p_{\rm c}$ is the probability for an X-ray source (with a given flux) to have a counterpart in a predefined catalogue, no information on the specific vicinity of this X-ray source is included.

The expression for the mean amount of candidates contained inside an annulus $i$ bounded by the circles with the radii ${r_i, r_{i + 1}}$ is the following:
\begin{equation}
\label{eqn:m2_m1}
   \lambda_{i}(r_i, r_{i + 1}; \sigma,p_{\rm c},\rho) = N_{\rm o}(r_{i + 1}) - N_{\rm o}(r_i).
\end{equation}

The probability function for the Poisson distribution:

\begin{equation}
    \Pr(\lambda_i, k_i) = \frac{e^{-\lambda_{i}} \lambda_{i}^{k_i}}{k_i!}.
\end{equation}
and the corresponding likelihood function:
\begin{equation}
    \La = \sum_{i} k_i \ln{\lambda_i} - \lambda_i,
    \label{eq:La}
\end{equation}
where $k_i$ is the observed amount of the optical candidates in the annulus $i$. We found parameters $\sigma,\ p_{\rm c},\ \rho$ of the model (\ref{eq:statistic}) through the maximum likelihood function (\ref{eq:La}) optimisation. The optimisation was performed in predefined bins in detection likelihood $\Lb$ and effective magnitude $m_{\rm eff}$. To take into account the photometric information, we built a family of the cross-match models for the optical objects filtered with the different threshold effective magnitude $m_{\rm eff}$. These threshold values were chosen basing on the quantiles of the optical counterparts distribution in $m_{\rm eff}$ for every detection likelihood range.

\begin{figure}
    \centering
    \includegraphics[width=\linewidth]{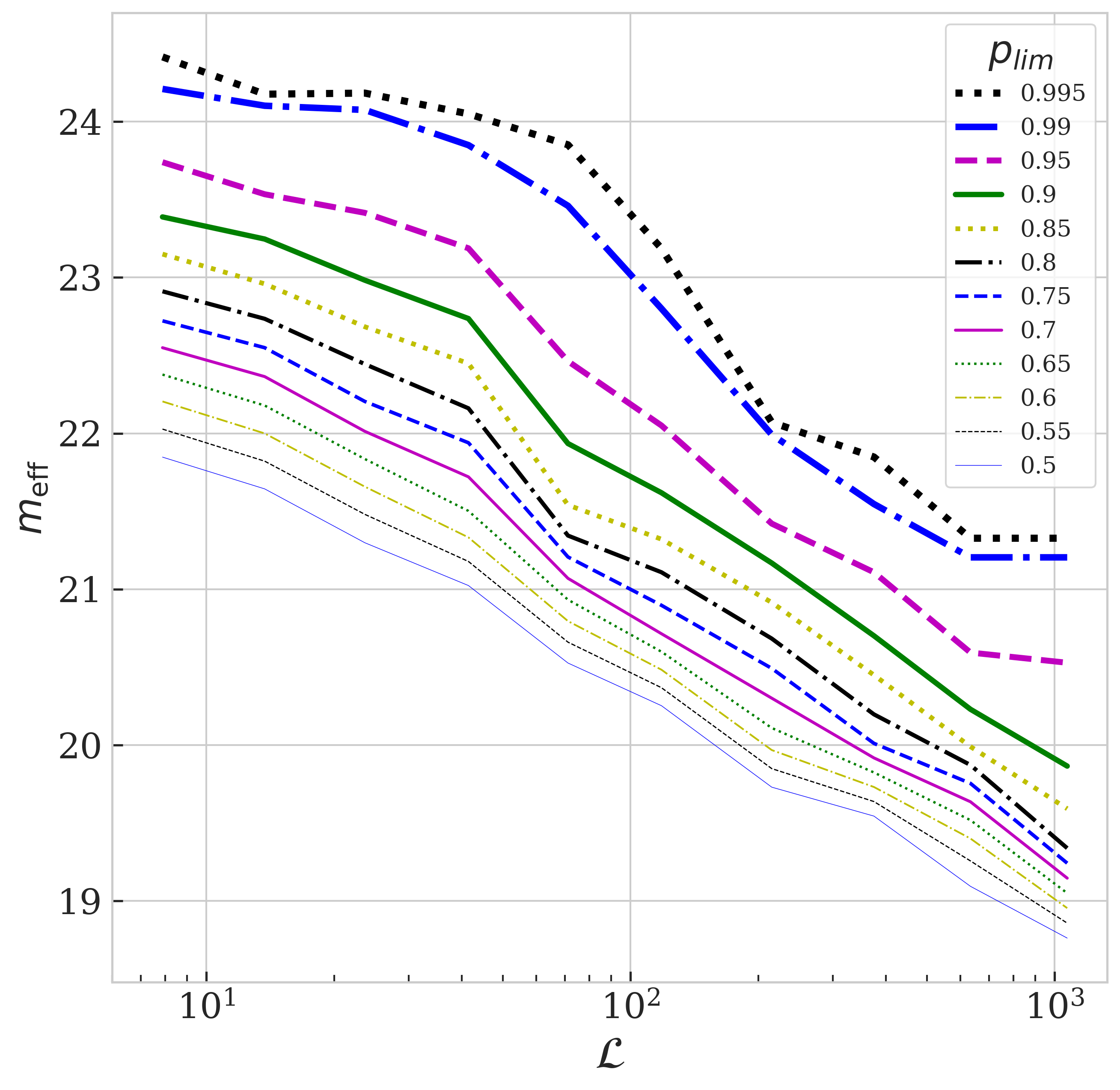}
    \begin{center}
    \caption{An effective magnitude vs detection likelihood $m_{\rm eff}(\Lb)$. The different lines correspond to the different threshold quantiles of the optical counterparts distribution in the effective magnitude.}
    \label{fig:dmag_plim}
    \end{center}
\end{figure}

\subsection{The relations between the model parameters and the detection likelihood/X-ray flux.}

To illustrate our method, in this Subsection we show how our cross-match model approximates the data for the effective magnitude threshold value chosen for \mbox{$p_{\rm lim}=0.85$} (the 85th percentile of the counterparts' distribution in $m_{\rm eff}$; see Fig \ref{fig:dmag_thresh}, vertical lines on the lower panel). In every detection likelihood range we define some unique value of $m_{\rm eff}$ chosen in accordance with the given $p_{\rm lim}$ ($0.85$ in this case). We also describe parameters behaviour for the different values of the $p_{\rm lim}$.

For \mbox{$p_{\rm lim}=0.85$} Fig.~\ref{fig:fit_in_bin} shows the optical objects distribution in the angular distance (within the circle $R<\ang{;;30}$) between the X-ray sources and the optical candidates. The distributions are shown in the different detection likelihood ($\Lb$) bins. As one can see the best models fit the data quite well. Additional information is presented in the Table \ref{table:pars_table}.

Fig.~\ref{fig:dmag_plim} shows the relations between the effective magnitude and the detection likelihood for various quantiles of the optical counterparts distribution. For a fixed $p_{\rm lim}$, $m_{\rm eff}(\Lb)$ is a monotonically decreasing dependence. The curves for the $p_{\rm lim}=50$\% and $99.5$\% quantiles differ by a factor of $\sim4-5$ in the surface density of the optical objects.

The \srge exposure time does not change much in the LH, so we can use the following expression to switch from the detection likelihood $\Lb$ to the X-ray flux in 0.5--2 keV range: $\lg(F_{\rm X,0.5-2}) = 0.66 \times \lg(\Lb) -15.09$. This expression approximates well the median values of the X-ray flux in the detection likelihood bins.

\begin{figure}
    \centering
    \includegraphics[width=\linewidth]{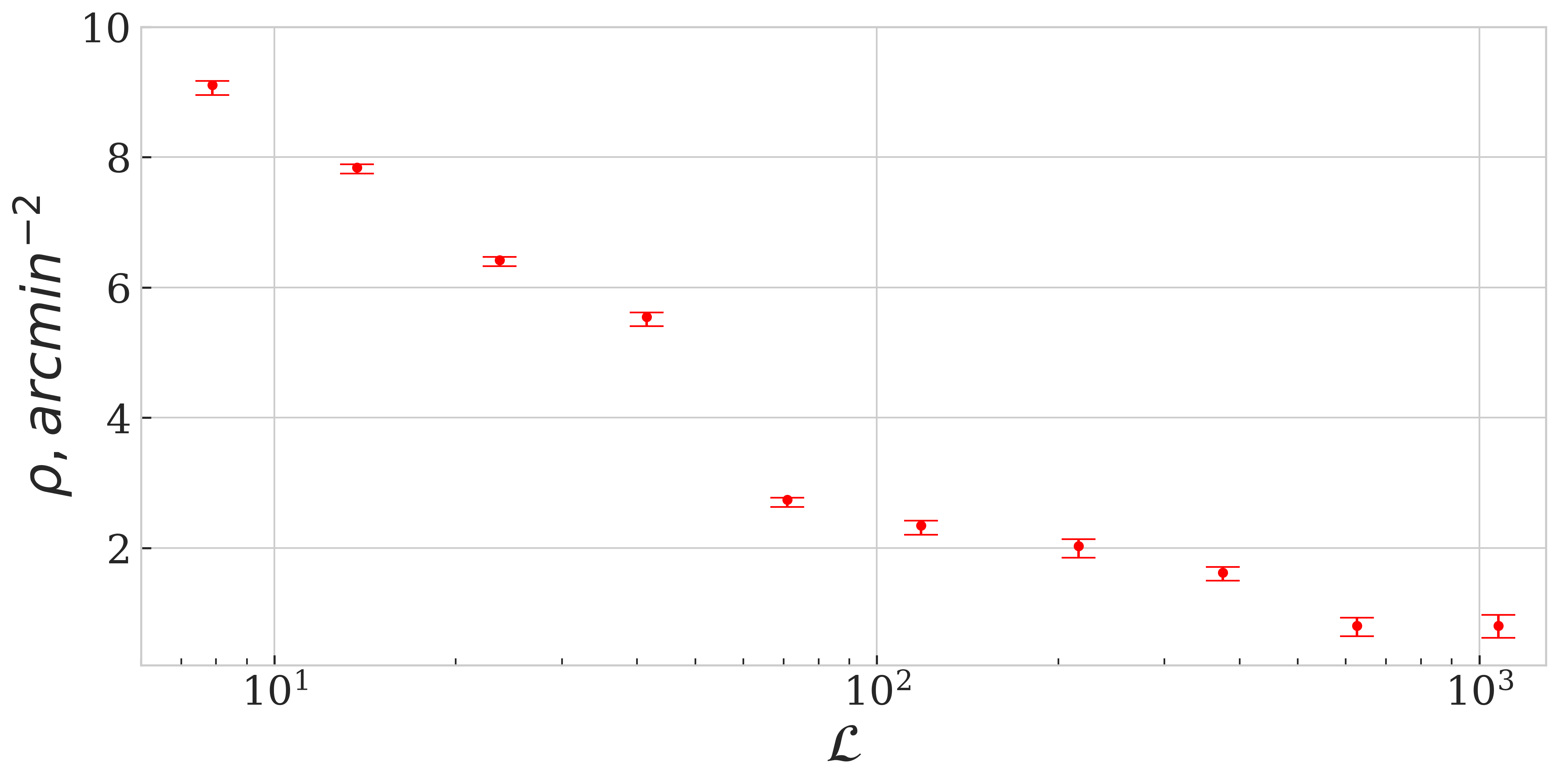}
    \begin{center}
    \caption{The surface density of the optical objects for the different detection likelihood ($\Lb$) bins; $p_{\rm lim} = 0.85$, (see \S\ref{chap:plim_model}).}
    \label{fig:rho_dl}
    \end{center}
\end{figure}

\begin{figure*}
    \centering
    \includegraphics[width=\linewidth]{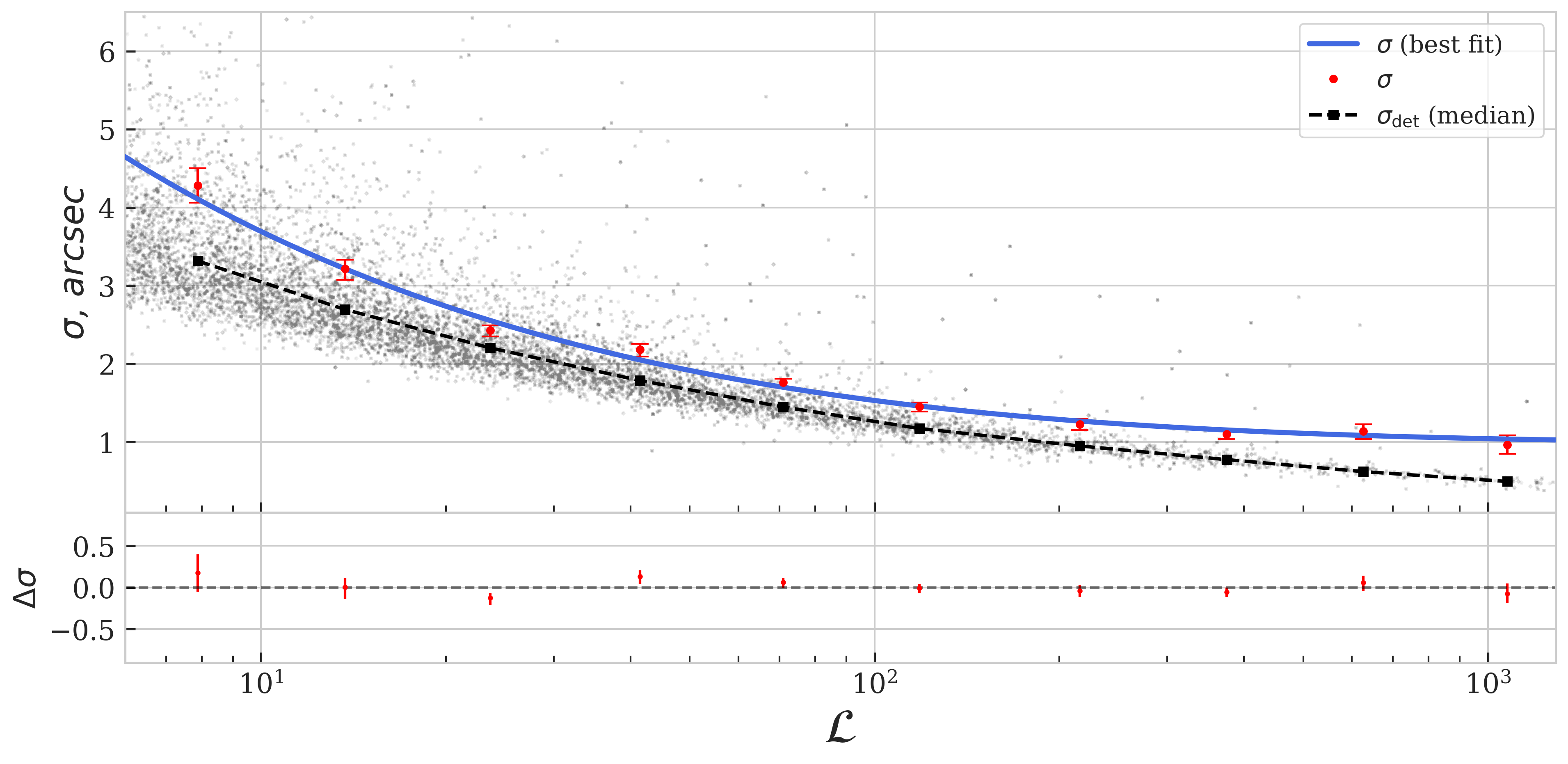}
    \begin{center}
    \caption{The dependency between the localisation error of the X-ray LH sources and their detection likelihood. The solid line ($\sigma$ best fit) is the best parametric approximation of the  $\sigma(\Lb)$ dependency; its residuals are shown in the lower panel. The scatter plot is the localisation errors $\sigma_{\rm det}$, calculated using the source-detection algorithm. The squares ($\sigma_{\rm det}$ median) mark their median values in the detection likelihood bins.}
    \label{fig:sigma_dl}
    \end{center}
\end{figure*}

Figs \ref{fig:rho_dl}, \ref{fig:sigma_dl} and \ref{fig:pc_flux} show the best values of the model parameters versus the detection likelihood (for $\rho$, $\sigma$) or the X-ray flux (for $p_{\rm c}$). We measured errors for the $\rho$, $\sigma$, $p_{\rm c}$ in each detection likelihood range using the likelihood ratio test (for example see \citealt{2006smep.book.....J}). We vary the value of a chosen parameter in the vicinity of the likelihood optimal value ($\La^{\rm best}$) and calculate the maximum likelihood for other parameters. 68\% confidence intervals for the chosen parameter match likelihood $\La^{\rm best} - \chi_{\text{2dof}}^{2}$.

\begin{figure} 
    \centering
    \includegraphics[width=\linewidth]{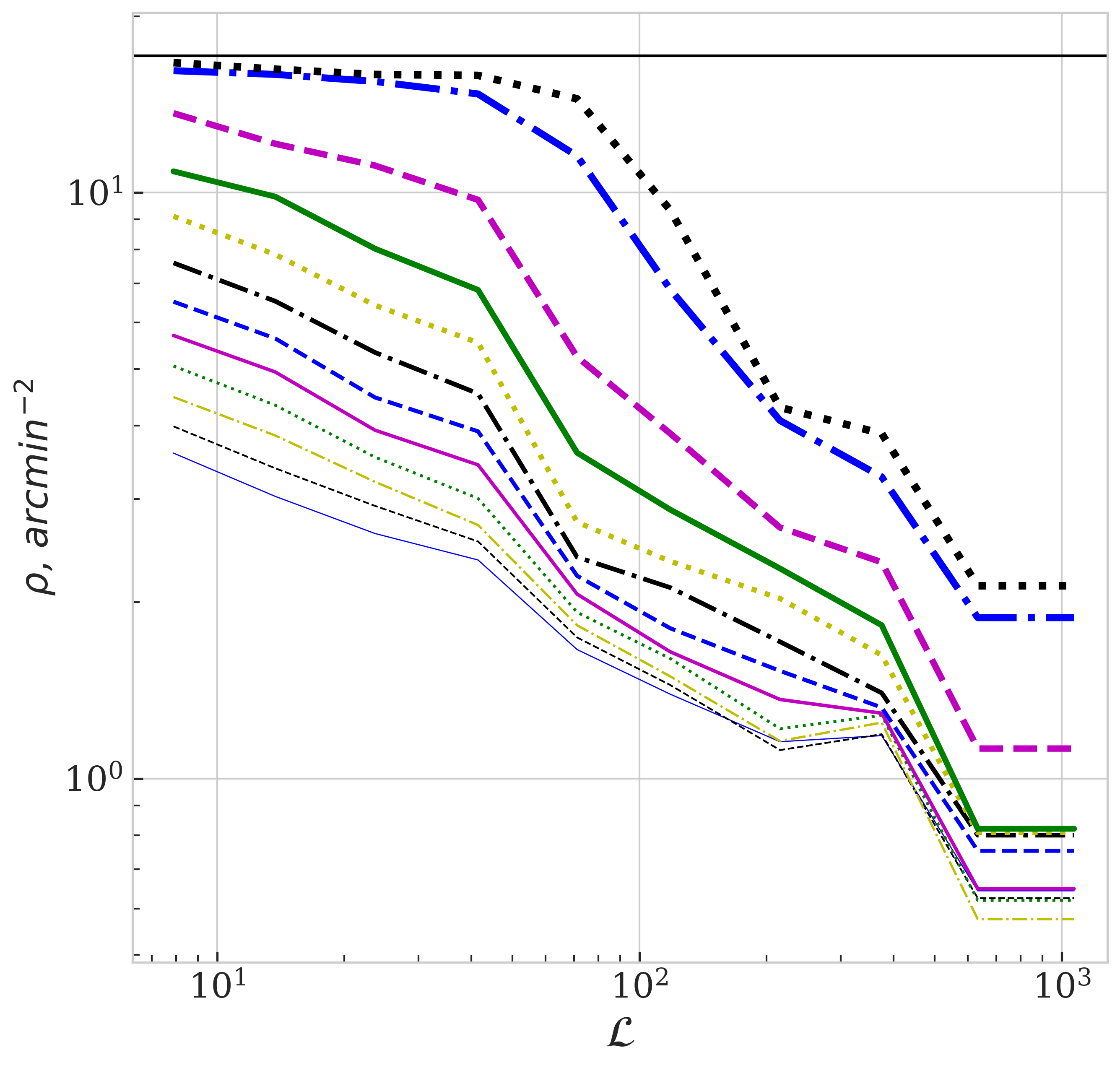}
    \begin{center}
    \caption{The relations between the surface density of the optical objects and the detection likelihood --- $\rho(\Lb)$. The curves are shown for the different quantiles ($p_{\rm lim}$) of the optical counterparts distribution. The horizontal line is a surface density calculated for all of the DESI LIS objects without any magnitude filters. See the legend on Fig. \ref{fig:dmag_plim}}
    \label{fig:rho_plim}
    \end{center}
\end{figure}

\subsubsection{$\rho$ vs $(\Lb)$}
The relationship between the surface density of the optical objects (selected using the effective magnitude threshold) and the detection likelihood --- $\rho(\Lb)$ --- is shown on Fig.~\ref{fig:rho_dl} (for $p_{\rm lim}=0.85$). The hallmark here is that the surface density decreases on order as the detection likelihood increases from \mbox{$\Lb=6$} to \mbox{$\Lb\approx10^3$}. This happens because bright sources (on average) have bright counterparts. For every detection likelihood range we choose unique $m_{\rm eff}$ threshold (in this current example the $m_{\rm eff}$ threshold corresponds to the $p_{\rm lim}=85$\%).

Fig.~\ref{fig:rho_plim} shows the dependencies between the optical objects surface density and the detection likelihood for the different values of $p_{\rm lim}$ (quantiles of the counterparts distribution). Without a magnitude threshold, the density is independent of the detection likelihood (a horizontal line). The curves $\rho(\Lb)$ for $p_{\rm lim}=50$\% and $99.5$\% differ by a factor of $\sim4-5$ in the surface density. This also happens because we choose different $m_{\rm eff}$ thresholds for different detection likelihood bins.

\begin{figure}
    \centering
    \includegraphics[width=\linewidth]{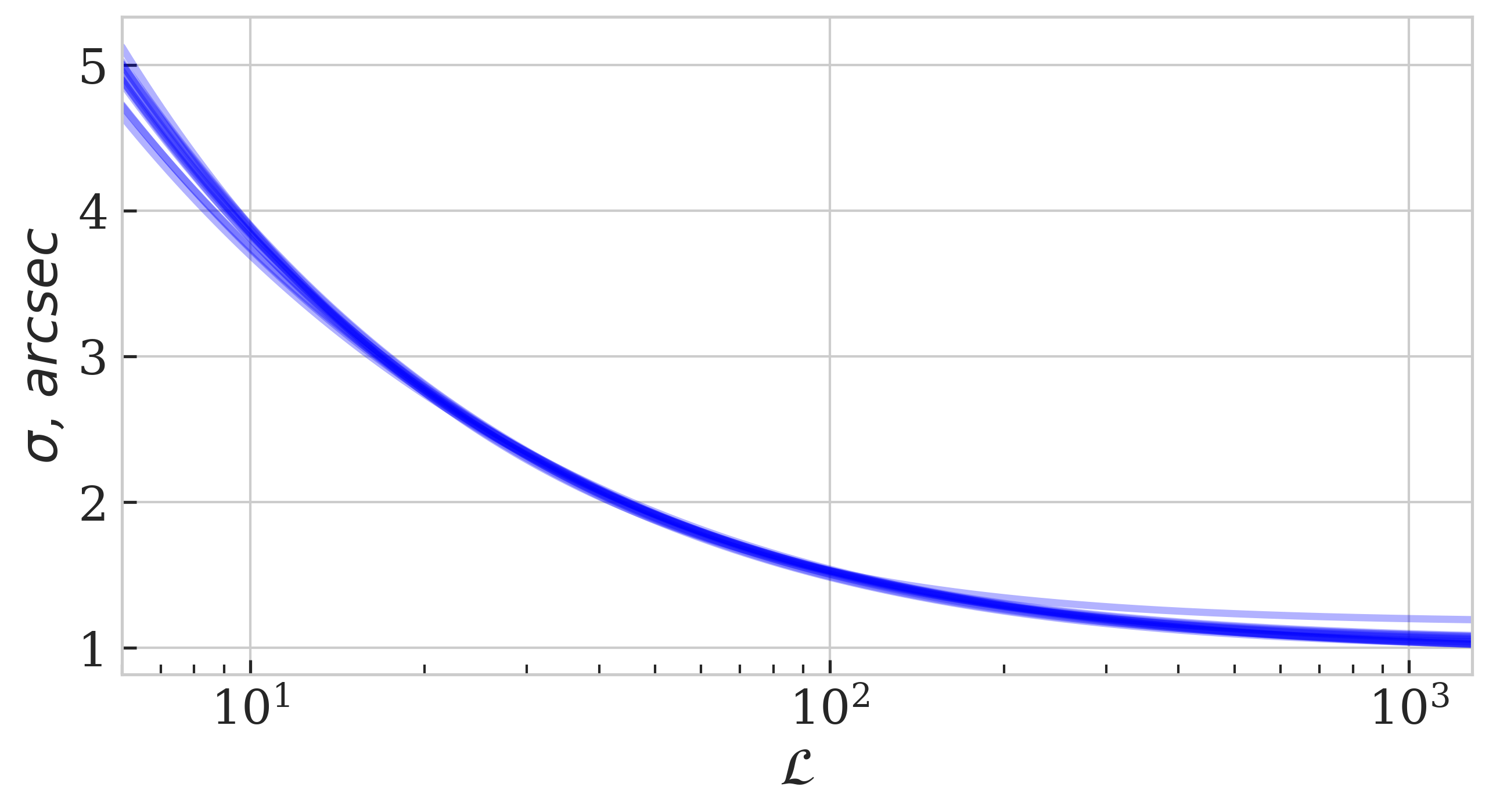}
    \begin{center}
    \caption{The dependency between an X-ray source localisation error and its detection likelihood --- $\sigma(\Lb)$. The dependencies are measured for the different $p_{\rm lim}$ values (counterparts distribution quantiles). The plot shows that the $\sigma(\Lb)$ behaviour is independent of the $p_{\rm lim}$.}
    \label{fig:sigma_plim}
    \end{center}
\end{figure}

\subsubsection{$\sigma$ vs $(\Lb)$}

Fig.~\ref{fig:sigma_dl} shows the dependency between the localisation error of the X-ray sources and their detection likelihood $\sigma(\Lb)$ (for $p_{\rm lim}=0.85$). The solid circles show the optimal $\sigma$ values found in the different $\Lb$ bins (with the corresponding errors). The scatter plot is the localisation errors $\sigma_{\rm det}$ (calculated using the detection algorithm) for all point-like X-ray LH sources. As one can see, the detection algorithm (on average) underestimates the true localisation error and needs to be calibrated using the optical data (see \S\ref{chap:sigma_result}).

The localisation errors obtained in this work for the different detection likelihood bins are well described by the following three-parameter dependence (for $p_{\rm lim}=0.85$; see Fig.~\ref{fig:sigma_dl}):

\begin{equation}
    \sigma (\mathcal{L}; a,b,\sigma_{0}) = \sqrt{a\mathcal{L}^b + \sigma_0^2},
\end{equation}
with the optimal parameters values $a=114.12\pm17.16,\ b=-0.96\pm0.04,\ \sigma_0=0.96\pm 0.12$.

Fig.~\ref{fig:sigma_plim} shows the best approximations for the dependency between the optical objects surface density and the detection likelihood built for the different $p_{\rm lim}$ values. The plot shows that the different $p_{\rm lim}$ values make almost no difference for the $\sigma(\Lb)$ dependency measurement.

\begin{figure}
    \centering
    \begin{center}
    \includegraphics[width=\linewidth]{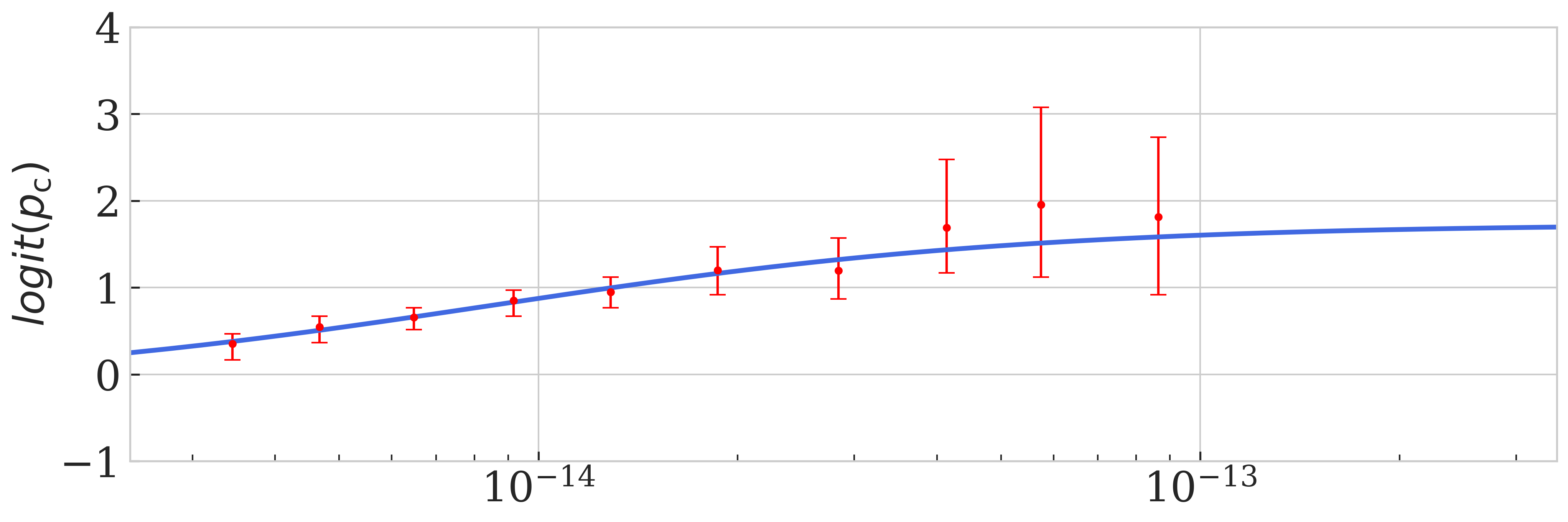}
    \includegraphics[width=\linewidth]{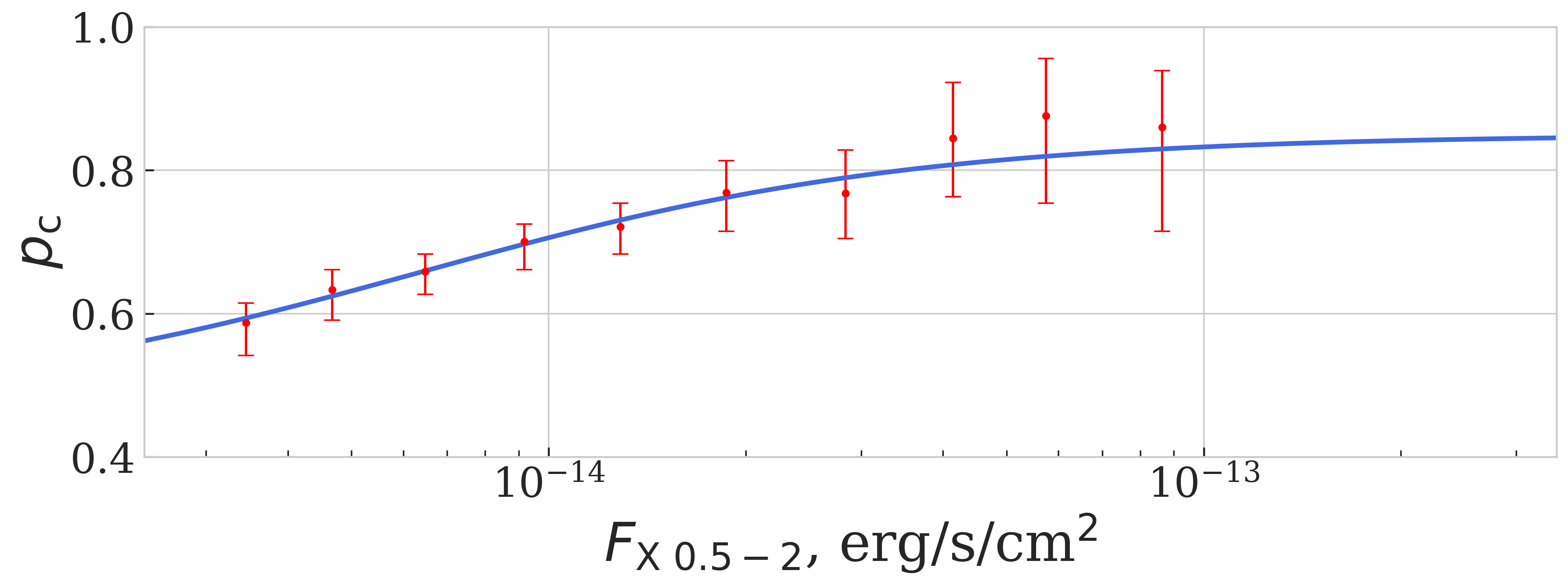}
    \caption{The dependency between the probability for an X-ray source to have an optical counterpart in the DESI LIS and its X-ray flux $F_{\rm X, 0.5 - 2}$ (for $p_{\rm lim} = 0.85$). The upper panel shows the $\text{logit}(p_{\rm c})$ values, the lower --- $p_{\rm c}$.}
    \label{fig:pc_flux}
    \end{center}
\end{figure}

\begin{figure}
    \centering
    \includegraphics[width=\linewidth]{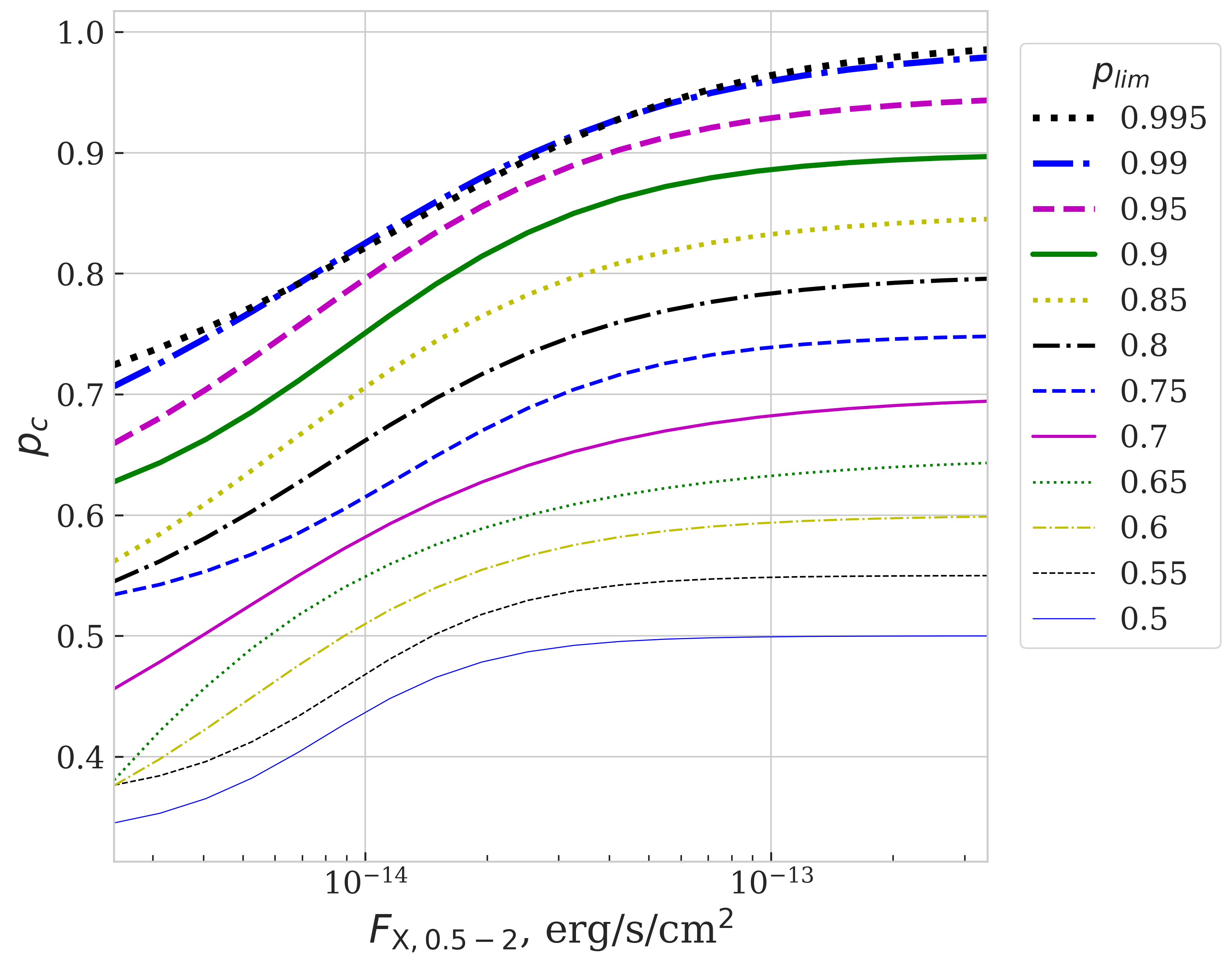}
    \begin{center}
    \caption{The probability for the \srge source to have an optical counterpart as a function of the X-ray flux --- $p_{\rm c}(F_{\rm X, 0.5 - 2})$. Different curves correspond to the different $p_{\rm lim}$ values (quantiles of the optical counterparts distribution in $m_{\rm eff}$).}
    \label{fig:pc_plim}
    \end{center}
\end{figure}

\subsubsection{$p_{\rm c}$ vs $F_{0.5-2}$}

Fig.~\ref{fig:pc_flux} (lower panel) shows the dependency between the probability for an X-ray source to have an optical counterpart and an X-ray flux $p_{\rm c}(F_{\rm X,0.5-2})$ for $p_{\rm lim}=0.85$. Dots are the measured $p_{\rm c}$ values for the corresponding flux bins.

The errors calculated for $p_{\rm c}$ are not Gaussian because the probability $p_{\rm c}$ lie within the interval $[0,1]$. To approximate the data with the parametric dependence we modified all calculated $p_{\rm c}$ values using logistic function

\begin{equation} \label{eqn:logit}
    \text{logit}(\rm p_{\rm c}) = \ln{\Big(\frac{p_{\rm c}}{1 - p_{\rm c}}\Big)} ~,
\end{equation}
so these transformed $p_{\rm c}$ values could lie within the whole number line ($-\infty,+\infty$).

Fig.~\ref{fig:pc_flux} (upper panel) shows the dependency between the value $\text{logit}(\rm p_{\rm c})$ and the X-ray flux. The errors for this value are close to the Gaussian. We approximate the dependency $\text{logit}(\rm p_{\rm c})$--- $\log{}F_{\rm X,0.5-2}$ using the following model:

\begin{equation} \label{eqn:sigmoid}
    \text{logit}(p_{\rm c}) = p_1 + \frac{\text{logit}(p_{\rm lim}) - p_1}{1 + \exp({-p_3 [\lg(F_{\rm X,0.5-2}) - p_2]})},
\end{equation}
where $p_1$, $p_2$, $p_3$ are the parameters of the logistic function which was obtained using the chi-squared minimization.

The solid line on Fig.~\ref{fig:pc_flux} (on both panels) is the best approximation of the data with model (\ref{eqn:sigmoid}). Note that to go back from $\text{logit}(\rm p_{\rm c})$ to $p_{\rm c}$ we use an inverse logistic function

\begin{equation}
    p_{\rm c} = \frac{\exp(\text{logit}(p_{\rm c}))}{1 + \exp(\text{logit}(p_{\rm c}))}.
    \label{eq:inv_logit}
\end{equation}

Fig.~\ref{fig:pc_plim} shows the best approximations of the probability for an X-ray source to have an optical counterpart depending on the X-ray flux $F_{\rm X,0.5-2}$. Different curves correspond to different $p_{\rm lim}$ values. One can see that as $p_{\rm lim}$ decreases, only the brightest candidates remain selected (and $p_{\rm c}$ also decreases).

\subsection{Positional-photometric model} \label{chap:plim_model}

Previously we showed how we build the positional cross-match models for the \srge sources using the effective magnitude ($m_{\rm eff}$) concept. $m_{\rm eff}$ depends on $p_{\rm lim}$ (chosen threshold quantile of the optical counterparts distribution).

To take into account the photometric information, we created a set of 12 positional cross-match models for the set of $p_{\rm lim}$ values\footnote{We chose the following $p_{\rm lim}$ values: $0.5$, $0.55$, $0.6$, $0.65$, $0.7$, $0.75$, $0.8$, $0.85$, $0.9$, $0.95$, $0.99$, $0.995$} and corresponding $m_{\rm eff}$ values. Each $m_{\rm eff}$ value for the set of models can be represented graphically in a magnitude space as a three-dimensional figure (see the shape constrained by the dashed lines on Fig.~\ref{dmag_cube}). The relationships between $m_{\rm eff}$, model parameters and detection likelihood (or X-ray flux) for all 12 $p_{\rm lim}$ are shown on Figs \ref{fig:dmag_plim}, \ref{fig:rho_plim}, \ref{fig:sigma_plim}, \ref{fig:pc_plim}. 

The family of these cross-match models (calculated for different $p_{\rm lim}$ values) together with the model built using all optical data (without any $m_{\rm eff}$ filters) we refer to as a positional-photometric cross-match model.

\subsection{$p_{\rm match}$ and $p_{\varnothing}$} \label{chap:pmatch_p0}

In this Subsection we explain how we calculate the probability for an X-ray object with a given flux and a given optical vicinity to have a counterpart in a predefined optical survey. We also introduce a probability $p_{\rm match}$ for an optical candidate to be a counterpart.

The probability for a given optical candidate $\text{i}$ at a distance $r_{\rm i}$ from an X-ray source (with parameters $\sigma$ and $p_{\rm c, i}^{\rm eff}$ and the surface density of the optical sources $\rho^{\rm eff}_{\rm i}$) to be a counterpart is described by the following expression:

\begin{equation}
    p_{\rm match} = \frac{p_{\rm c, i}^{\rm eff} \exp \Big(\frac{-r_{\rm i}^2}{2 \sigma^2}\Big)}{p_{\rm c, i}^{\rm eff} \exp \Big(\frac{-r_{\rm i}^2}{2 \sigma^2}\Big) + 2 \pi \rho^{\rm eff}_{\rm i} \sigma^2} ~.
\end{equation}

Values $p_{\rm c, i}^{\rm eff},\ \rho^{\rm eff}_{\rm i}$ used here are obtained with the positional-photometric model (see \S\ref{chap:plim_model}). For a given candidate we choose a model from the set of the positional models with different $p_{\rm lim}$ values. We choose it in such a way that the model's $m_{\rm eff}$ appears to be as close as possible (but greater) to the candidate's $m_{\rm eff}$. For the candidates with the effective magnitude higher than the maximum $m_{\rm eff}$ available in the positional-photometric model (corresponding to $p_{\rm lim} = 0.995$) we use the model built using all optical data (without any $m_{\rm eff}$ filters): $p_{\rm c}^{\rm pos},\ \rho^{\rm pos}$.

The probability for an X-ray source with a given optical vicinity \textit{not to have} an optical counterpart we define as the following:
\begin{equation}
    p_{\varnothing} = \frac{1}{1 + \frac{1}{2 \pi \sigma} \sum_{i=1}^{n_{\rm o}}
    \exp\Big({ \frac{-r_{\rm i}^{2}}{2 \sigma^2}}\Big) \frac{p'_{\rm c, i}}{\rho'_{\rm i}}},
\end{equation}

where $n_{\rm o}$ is a number of the DESI LIS optical candidates in the vicinity of an \srge source (within \ang{;;30} in our case); $p'_{\rm c, i} = p_{\rm c}^{\rm pos} - \overline{p_{\rm c, i}^{\rm eff}}$, $\rho'_{\rm i} = \rho^{\rm pos} - \overline{\rho^{\rm eff}_{\rm i}}$. Parameters $\overline{p_{\rm c, i}^{\rm eff}},\ \overline{\rho^{\rm eff}_{\rm i}}$ as well as $p_{\rm c, i}^{\rm eff},\ \rho^{\rm eff}_{\rm i}$ are calculated using the model with $m_{\rm eff}$ as close as possible to $m_{\rm eff}$ of a candidate. Unlike with $p_{\rm match}$, model's $m_{\rm eff}$ should be lower than the canditate's $m_{\rm eff}$.

\subsection{Metrics for assessing the quality of the cross-match and source selection} \label{chap:metrics}

To evaluate the overall precision of our cross-match model we use the following metric:

\begin{equation}
    \text{Precision}_{\rm X} = \frac{\hat{N}^{\star}_{\rm c+h}}{N_{\rm X}} ~,
\end{equation}
where $N_{\rm X}$ is a total number of all X-ray sources in a sample, $\hat{N}^{\star}_{\rm c+h}=\hat{N}^{\star}_{\rm c} + \hat{N}^{\star}_{\rm h}$ is a number of X-ray sources for which we found a correct optical match ($\hat{N}^{\star}_{\rm c}$) or figured out (correctly) that there are no counterparts in a given photometric survey ($\hat{N}^{\star}_{\rm h}$).

To evaluate the quality of the counterpart selection process we use the following metrics:

\begin{equation}
\text{Recall}_{\rm c} = \frac{ \hat{N}^\star_{\rm c} }{ N_{\rm c} } ~,
\end{equation}
where $N_{\rm c}$ is a total number of all X-ray sources with counterparts in a sample.

\begin{equation}
\text{Precision}_{\rm c} = \frac{\hat{N}^\star_{\rm c} }{\hat{N}_{\rm c}} ~,
\end{equation}
where $\hat{N}_{\rm c}$ is a number of X-ray sources selected by our algorithm as sources with counterparts.

To evaluate the quality of selection for sources without counterparts (hostless) in the DESI LIS we use the following metrics:

\begin{equation}
\text{Recall}_{\rm h} = \frac{ \hat{N}^\star_{\rm h} }{ N_{\rm h} } ~,
\end{equation}
where $N_{\rm h}$ is a total number of X-ray sources without optical counterparts in a sample ($h$ is for \textit{hostless}).

\begin{equation}
\text{Precision}_{\rm h} = \frac{\hat{N}^\star_{\rm h} }{\hat{N}_{\rm h}} ~,
\end{equation}
where $\hat{N}_{\rm h}$ is a number of X-ray sources selected by our algorithm as sources without counterparts.

\section{Results and discussion} \label{results}

The main results of our work are the following.

In \S\ref{chap:sigma_result} we present the relationship $\sigma(\Lb)$ between the localisation error of the \srge LH X-ray sources and their detection likelihood. This dependence was measured using the information on optical objects surrounding the \srge LH sources. Analysing the relationship $\sigma(\Lb)$ and the averaged information on the localisation error $\sigma_{\rm det}(\Lb)$ we find a calibration relationship $\sigma(\sigma_{\rm det})$ and compare it with the results from the literature.

In \S\ref{chap:pc_result} we provide the relationship between the probability for an \srge LH X-ray source to have an optical counterpart (for the DESI LIS and for the SDSS) and its X-ray flux: $p_{\rm c}(F_{\rm X, 0.5 - 2})$.

In \S\ref{chap:posphot_result} we show how the counterpart searching can be improved with the photometric information on candidates.

\S\ref{chap:main_result} contains the information on the quality of optical cross-match we achieve for the whole \srge LH X-ray sample.

Finally, in \S,\ref{chap:selection_result} we provide results our algorithm allows achieving in searching optical counterparts and selecting X-ray sources without counterparts (hostless X-ray sources).

\begin{figure}
    \centering
    \includegraphics[width=\linewidth]{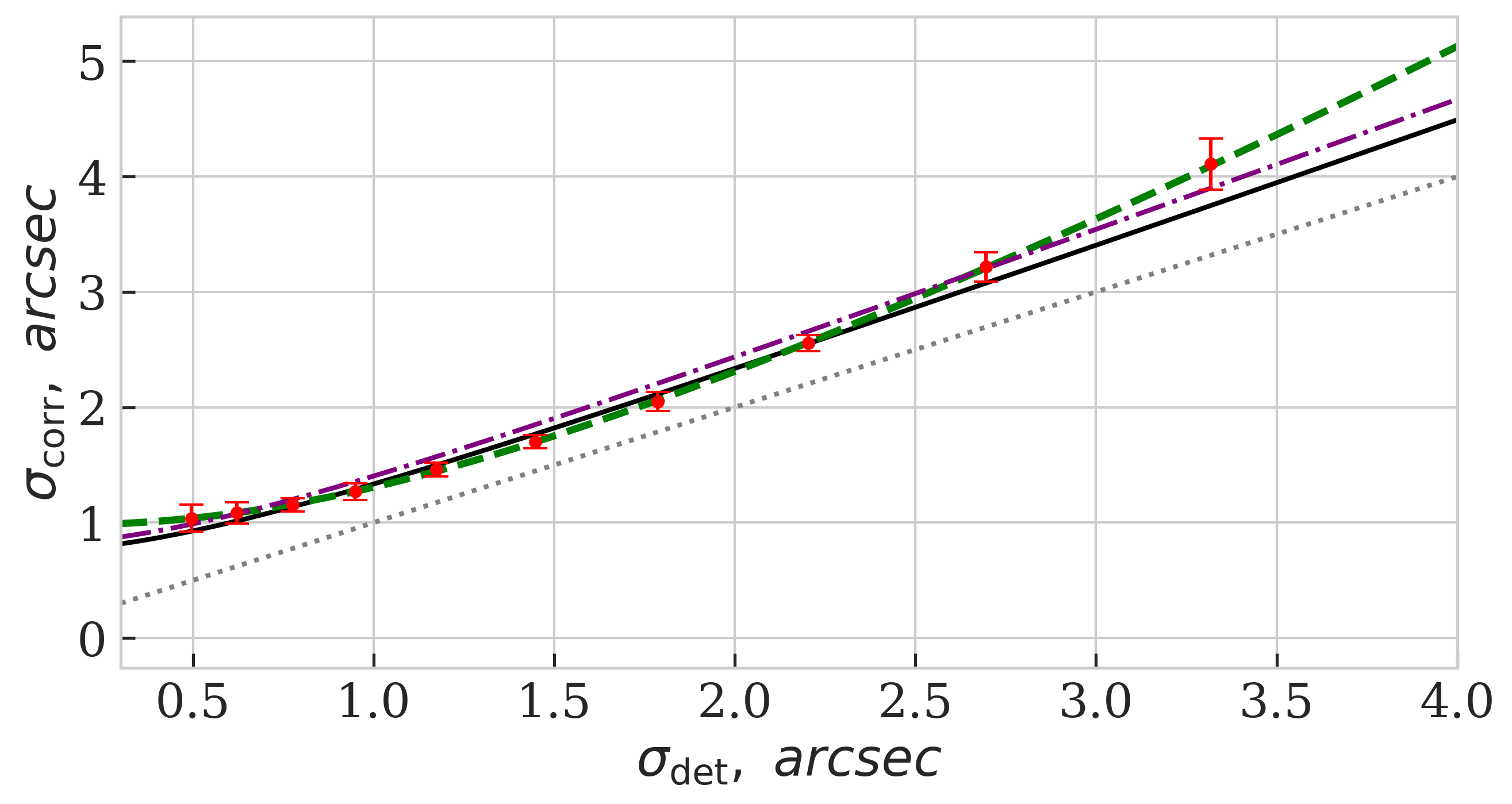}
    \begin{center}
    \caption{Calibration relationship between an average localisation error of point-like X-ray sources calculated 1) using the information on optical objects surrounding the \srge LH sources ($\sigma_{\rm corr}$) and 2) using the X-ray source detection algorithm eSASS ($\sigma_{\rm det}$). The solid line is a two-parameter model; the dashed line is a three-parameter model. The dash-dotted line is the relationship taken from \citealt{2021arXiv210614517B} (measured for another field eFEDS). The dotted line is $\sigma_{\rm corr}=\sigma_{\rm det}$.}
    \label{fig:esass_corr}
    \end{center}
\end{figure}

\subsection{Calibration of a localisation error of the \srge sources} \label{chap:sigma_result}

One of the main characteristics of the cross-match model is the relationship between a localisation error and detection likelihood for X-ray sources. We measured this relationship while searching for the best model parameters in different detection likelihood ($\Lb$) bins. We used the information obtained from the DESI LIS on optical candidates surrounding the \srge LH sources.

We found the calibration relationship between the localisation error of point-like X-ray sources calculated using the information on optical objects surrounding the \srge LH sources ($\sigma_{\rm corr}$) and using the X-ray source detection algorithm eSASS ($\sigma_{\rm det}$). We measured the median values of $\sigma_{\rm det}$ for X-ray sources in different detection likelihood bins (see the square markers connected with the dashed line on Fig.~\ref{fig:sigma_dl}). The calibration relationship $\sigma_{\rm corr}(\sigma_{\rm det})$ is shown on Fig.~\ref{fig:esass_corr}. The solid and dashed lines show the best ( $\chi^2$ optimisation) for the two- and three-parameter models, respectively.

\begin{equation}
    \sigma_{\rm corr,1} = a\sqrt{ \sigma_{\rm det}^2 + c^2} ~,
\label{eqn:sigma_corr_1}
\end{equation}

where $a = 1.11\pm0.03,\ c=0.68\pm0.07$;

\begin{equation}
    \sigma_{\rm corr,2} = a\sqrt{ \sigma_{\rm det}^{b} + c^2} ~,
\label{eqn:sigma_corr_2}
\end{equation}

where $a = 0.87\pm0.08,\ b=2.53\pm0.20,\ c=1.12\pm0.19$.

Both models describe the data well (taking into account the errors). The dotted line is $\sigma_{\rm corr}=\sigma_{\rm det}$. Note that on average a localisation error $\sigma_{\rm corr}$ we estimated is higher than $\sigma_{\rm det}$ (calculated using the eSASS algorithm).

The dash-dot line on Fig.~\ref{fig:esass_corr} is the calibration relationship taken from \citealt{2021arXiv210614517B} (measured using the eFEDS data).

\begin{equation}
    \sigma_{\rm corr,eFEDS} = 1.15\sqrt{\sigma_{\rm det}^2 + 0.7^2} ~.
\end{equation}

We can conclude that there is a good agreement between our calibration relationship $\sigma_{\rm corr}(\sigma_{\rm det})$ and the relationship presented in \citealt{2021arXiv210614517B}.

\begin{figure}
    \centering
    \begin{center}
    \includegraphics[width=\linewidth]{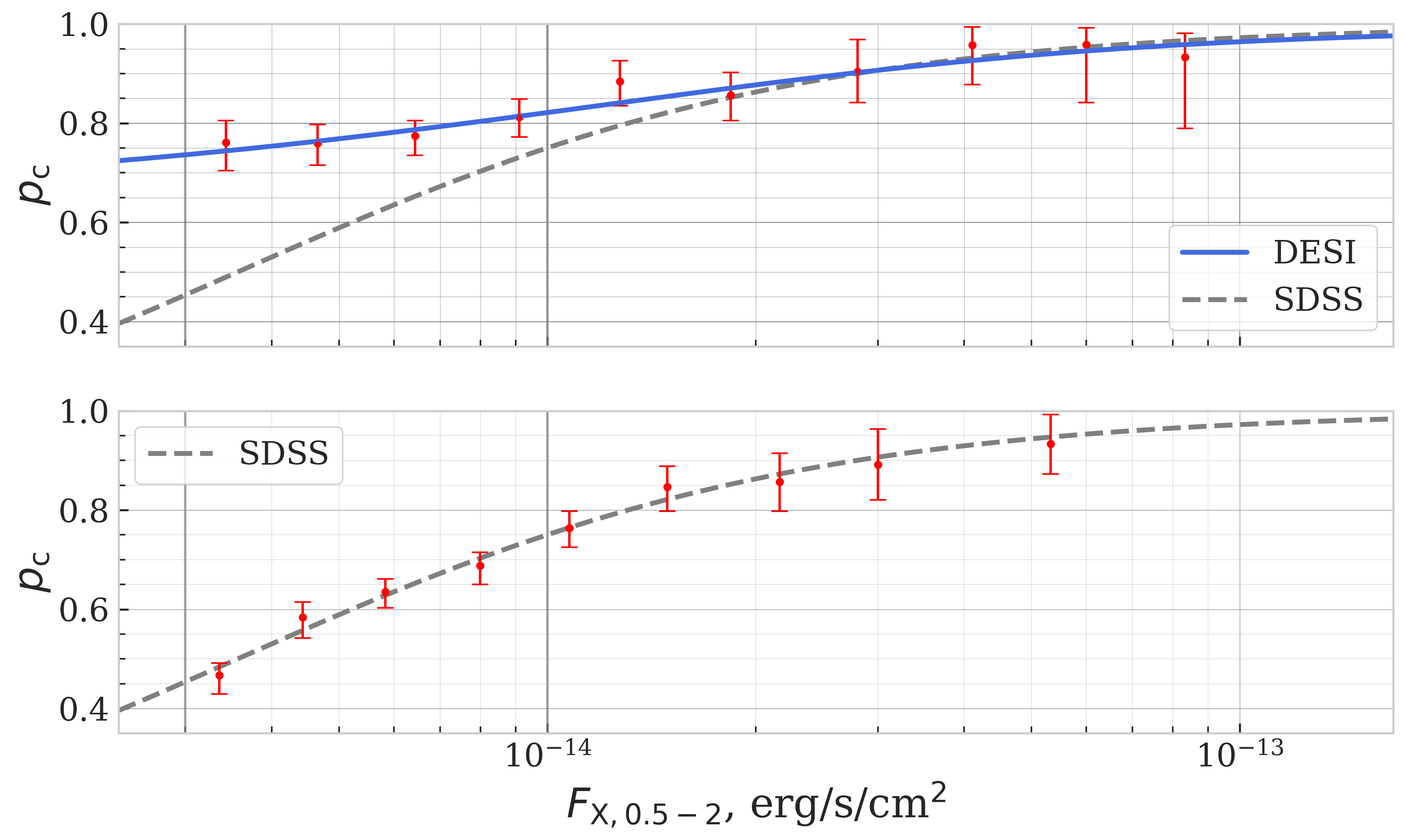}
    \caption{$p_{\rm c}$ vs $F_{\rm X, 0.5 - 2}$. The solid line is the relationship calculated for the DESI LIS, the dashed one is for SDSS. The vertical lines mark the threshold values for the flux ($F_{\rm X, 0.5 - 2}$) \num{3e-15} and $10^{-14}$ \xdim{}.}
    \label{fig:pc_flux_result}
    \end{center}
\end{figure}

\subsection{$p_{\rm c}$ vs $F_{\rm X, 0.5 - 2}$} \label{chap:pc_result}

It is essential to take into account that an optical counterpart may be absent in a photometric catalogue. Using the \srge Lockman Hole data, we measured a probability for an X-ray source with a given flux to have a counterpart in the DESI LIS or SDSS (Fig.~\ref{fig:pc_flux_result}, upper and lower panel accordingly). The best approximations (see equations \ref{eqn:sigmoid}, \ref{eq:inv_logit}) of the dependencies $p_{\rm c}(F_{\rm X, 0.5 - 2})$ we measured for the DESI LIS and SDSS are shown on Fig.~\ref{fig:pc_flux_result} by the solid and dashed lines, accordingly.

Fig.~\ref{fig:pc_flux_result} shows that a probability to have a counterpart $p_{\rm c}$ is $\approx74$\% and $\approx$45\% for an X-ray source with $F_{\rm X,0.5-2}\approx~\num{3e-15}$\,\xdim (for the DESI LIS and SDSS, respectively). For $F_{\rm X,0.5-2}\approx~10^{-14}$\,\xdim{} (corresponds to the expected four-year equatorial sensitivity for \srge) $p_{\rm c}$ values are similar both for the DESI LIS and SDSS: $\approx82$\% and $\approx75$\%. For $F_{\rm X,0.5-2}\gtrsim~10^{-14}$\,\xdim{} the fraction of X-ray objects having an optical counterpart becomes indistinguishable for both photometric surveys in question (within the errors).

\subsection{Optical cross-match using the effective magnitude} \label{chap:posphot_result}

\begin{figure}
    \centering
    \includegraphics[scale=0.4]{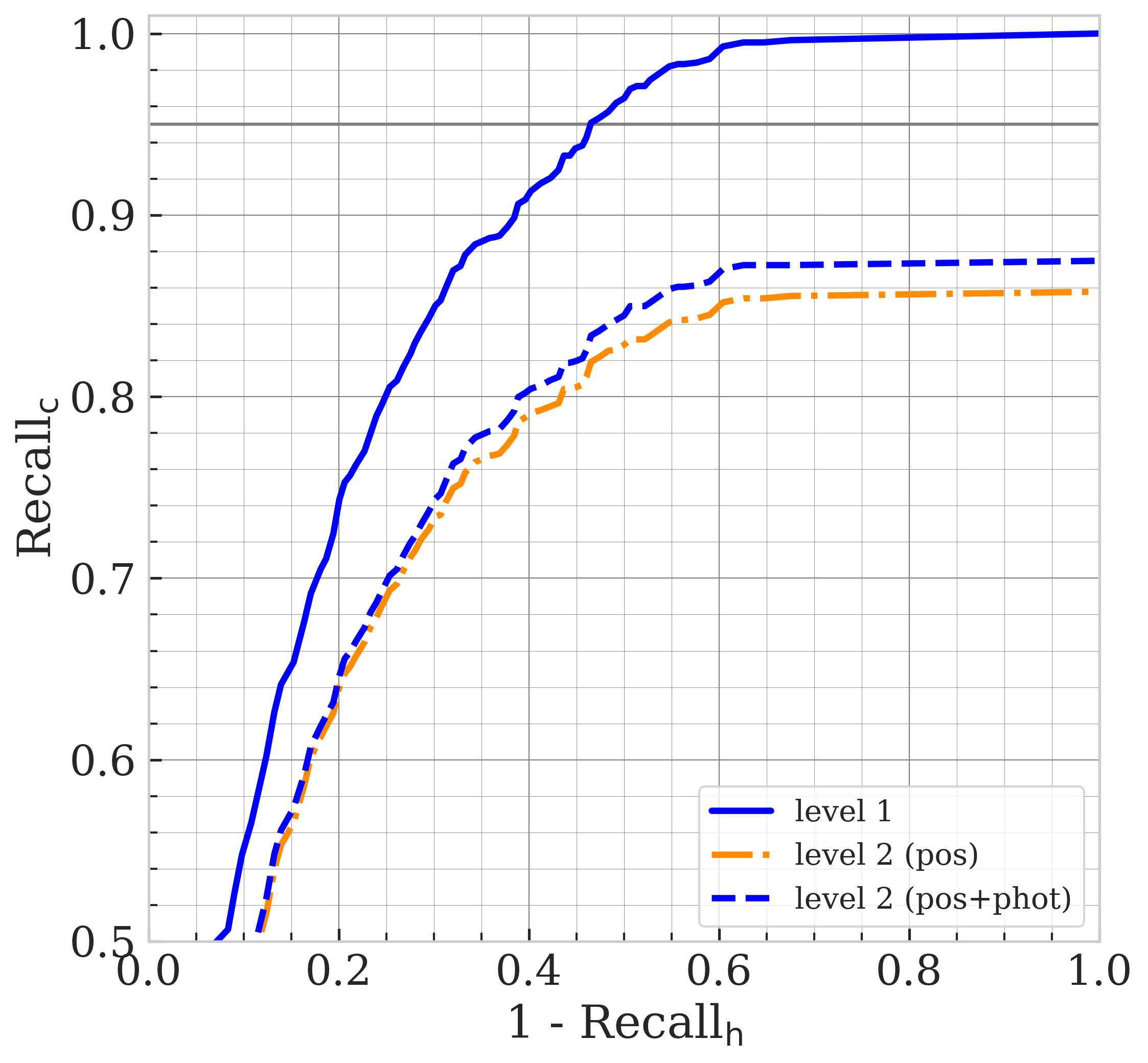}
    \begin{center}
    \caption{The relationships between the selection recall of X-ray sources with counterparts ($\text{Recall}_{\rm c}$) and the ratio of missed (during the selection process) fields \textit{without} optical counterparts ($\text{Recall}_{\rm h}$). The relationships are built for $F_{\rm X, 0.5 - 2} > \num{3e-15}$~\xdim{}. The solid line shows the result for level 1 of our cross-match model (see Fig.~\ref{fig:cross_scheme}). The dashed line is the result of the selection using both positional and photometric information (level 2 pos+phot). The dash-dotted line is the result of the selection based only on positional information (i.e. when choosing the closest optical candidate as a counterpart, level 2 pos).}
    \label{fig:roc_pseudo}
    \end{center}
\end{figure}

False optical candidates (objects appeared nearby the X-ray source by chance) and counterparts belong to different distributions both in separation and magnitude. We use photometric information on candidates in a form of the effective magnitude, $m_{\rm eff}$. This approach allows reducing the number of missed counterparts. Fig.~\ref{fig:roc_pseudo} shows the recall curves for three different types of X-ray sources selection: 1) selection of sources having a counterpart, no matter if it identified correctly or not (the solid line); 2) selection of sources having a counterpart identified using only positional information (the dash-dotted line) 3) selection of sources identified using both positional and photometric information (the dashed line). To draw the solid line we used optical fields containing counterparts as the positive class objects. For the two other lines, positive class objects are the fields where counterpart is present \textit{and identified} using positional information \textit{or} both positional and photometric information.

Fig.~\ref{fig:roc_pseudo} shows that for the DESI LIS photometric information allows reducing the number of counterparts missed during the selection process by $\approx$13\% (for X-ray sources with $F_{\rm X, 0.5 - 2} > \num{3e-15}$~\xdim{}).

\subsection{Overall cross-match quality for the whole LH sample} \label{chap:main_result}

\begin{figure*}
    \centering
    \begin{subfigure}{.4\linewidth}
      \centering
      \includegraphics[scale=.4]{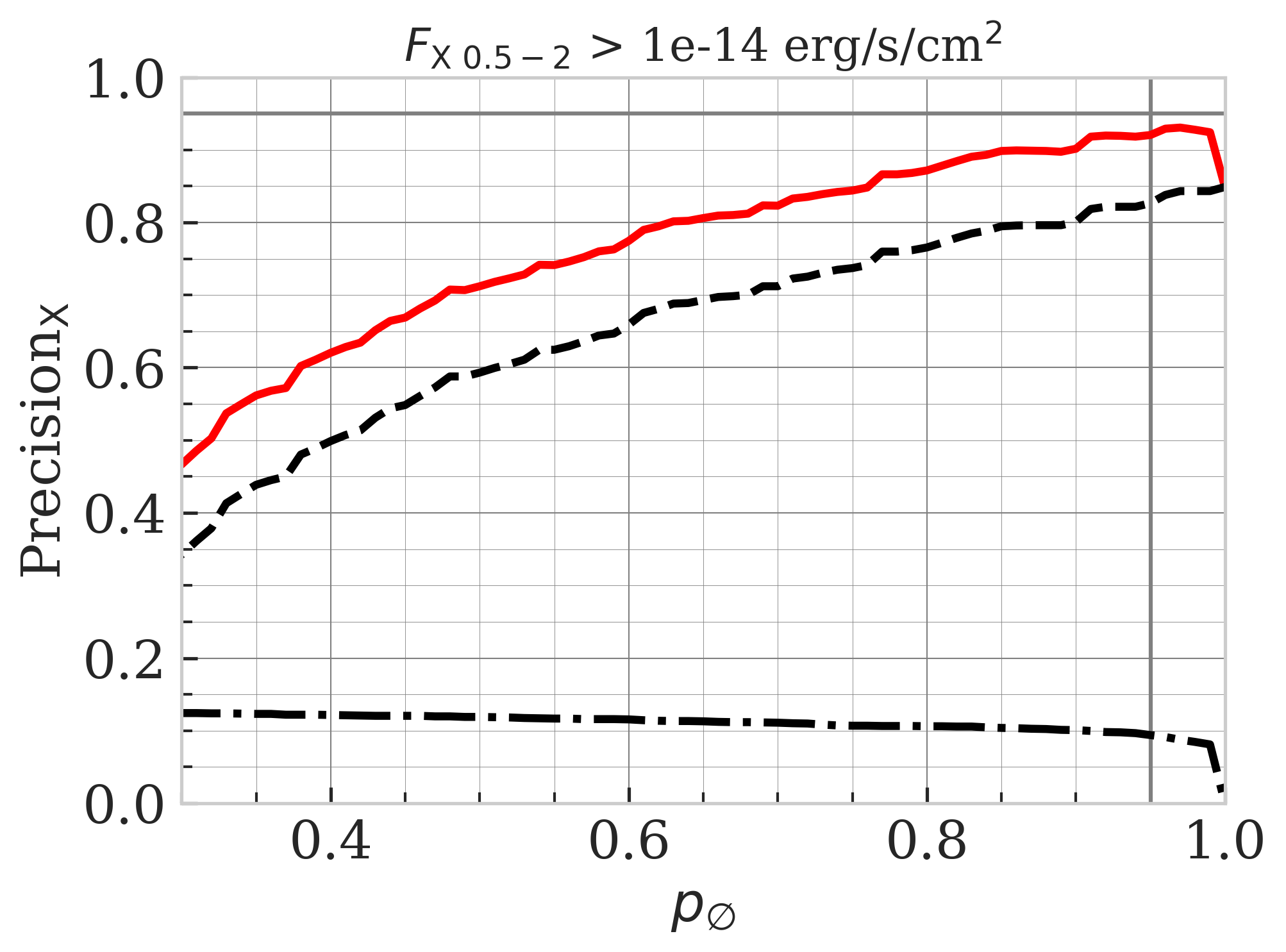}
    \end{subfigure}
    \begin{subfigure}{.25\linewidth}
      \centering
      \includegraphics[scale=.4]{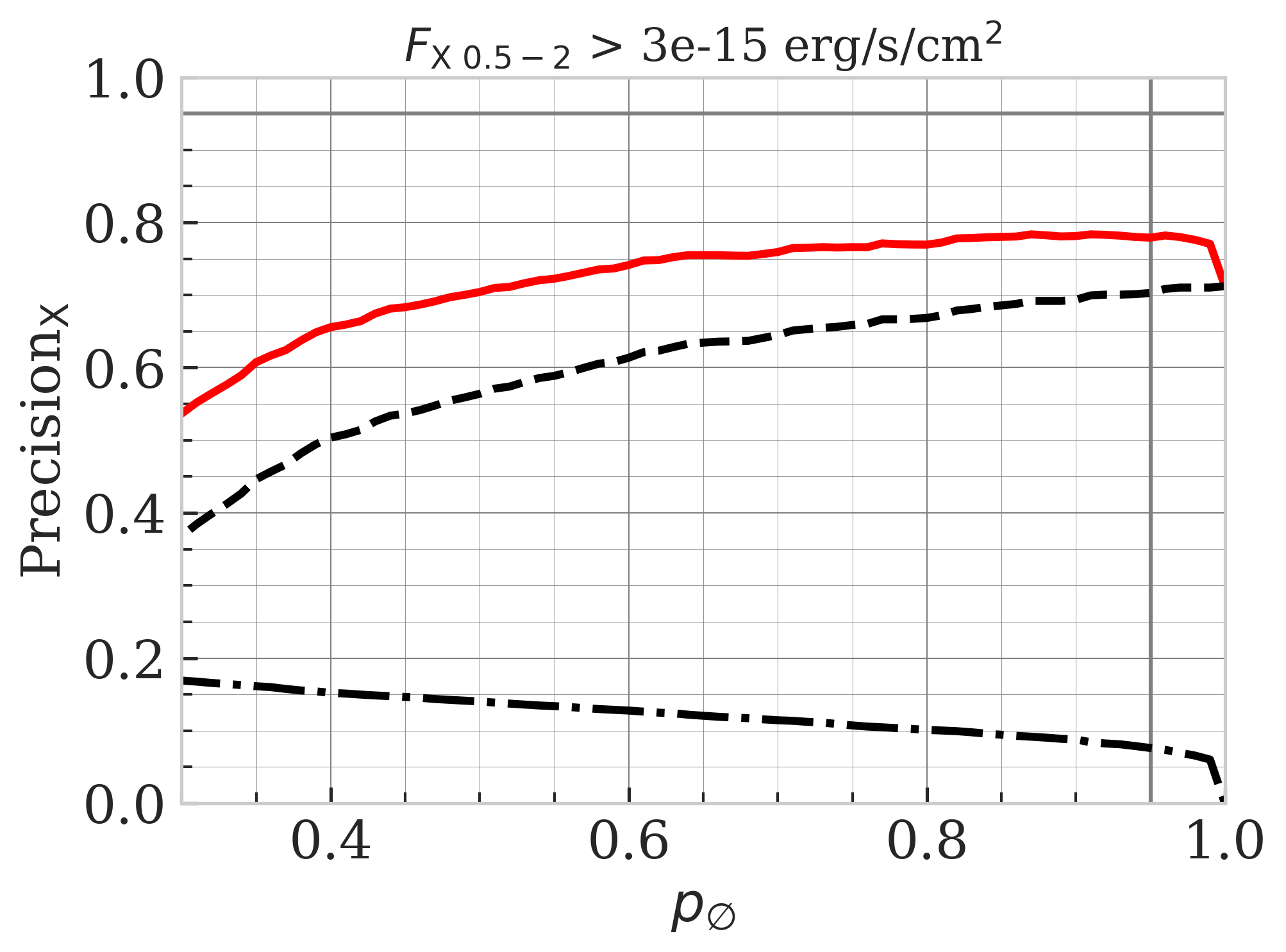}
    \end{subfigure}
    \caption{The overall precision (the solid line). To calculate the overall precision we define a correct match either as a correctly identified counterpart ($\hat{N}^{\star}_{\rm c} / N_{\rm X}$, the dashed line) or as a correctly revealed X-ray source without a counterpart ($\hat{N}^{\star}_{\rm h} / N_{\rm X}$, the dash-dotted line). The metrics were calculated for the DESI LIS for two different X-ray flux thresholds. The vertical line shows $p_{\varnothing} = 0.95$.}
    \label{fig:metrics_pseudo_total}
\end{figure*}

The overall precision of our cross-match procedure for the \srge LH sources with $F_{\rm X, 0.5 - 2} > 3\times 10^{-15}$~\xdim{} is 78\% (see Fig.~\ref{fig:metrics_pseudo_total}, right panel, the solid line). For sources with flux higher than $10^{-14}$ --- 93\% (see same figure, left panel, solid line). To calculate the overall precision we define a correct match either as a correctly identified counterpart or as a correctly revealed X-ray source without a counterpart. Precision and recall vary depending on the value of $p_{\varnothing}$ (the probability that there is no counterpart for an X-ray source in the DESI LIS \textit{considering the vicinity of this X-ray source}).

\subsection{Search of optical counterparts and selection of hostless X-ray sources} \label{chap:selection_result}

When performing a cross-mach, we need two different selection regimes: with a high precision (80\% and higher) or with high recall (90\% and higher). The former regime is preferable when we search for unique and rare sources; the latter suits the tasks similar to QSO luminosity function measurement (when one needs to reduce the selection effects). Our model allows working in both regimes by varying parameter $p_{\varnothing}$.

\begin{figure*}
    \centering
    \begin{subfigure}{.4\linewidth}
      \centering
      \includegraphics[scale=.4]{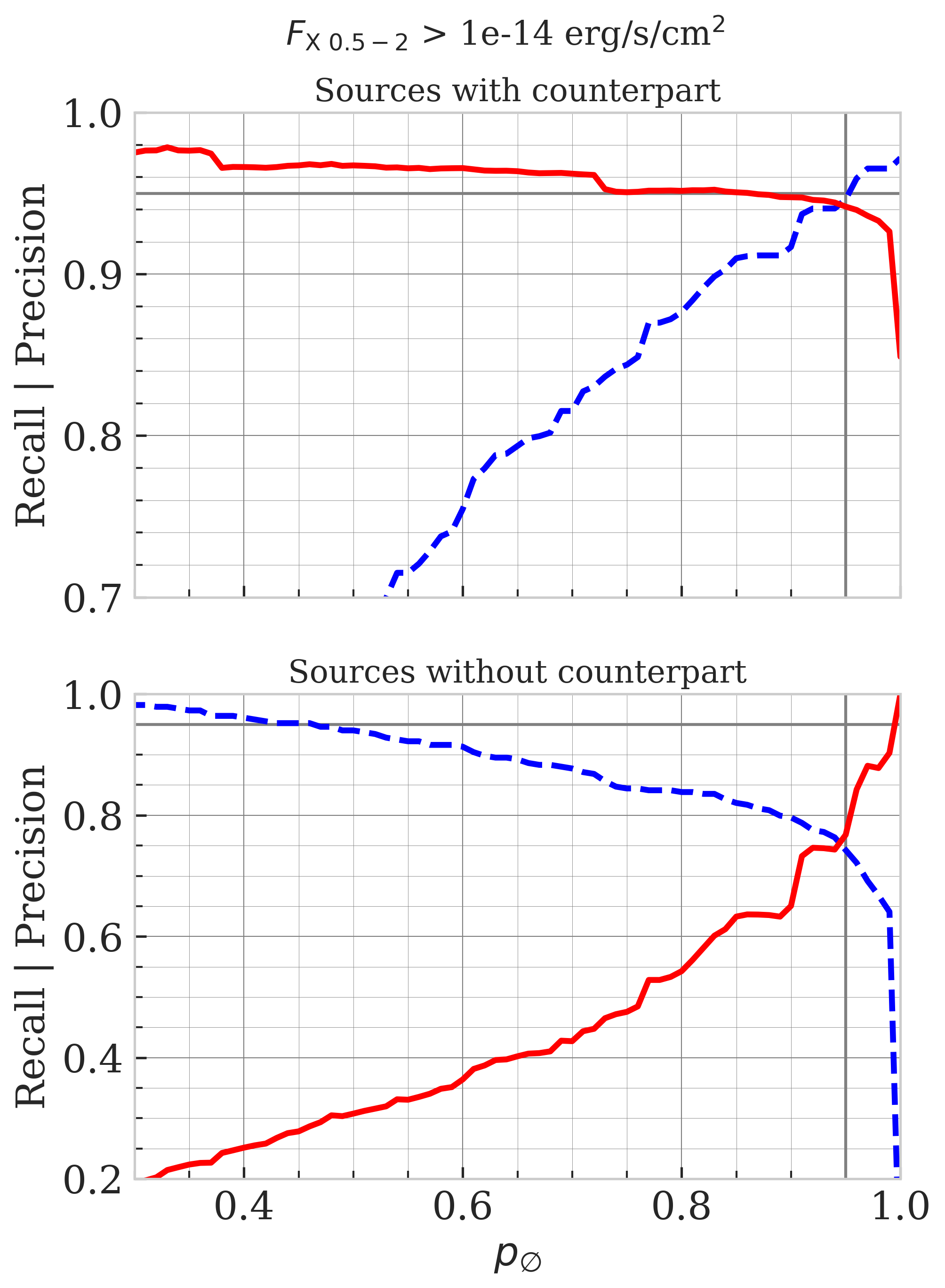}
    \end{subfigure}%
    \begin{subfigure}{.25\linewidth}
      \centering
      \includegraphics[scale=.4]{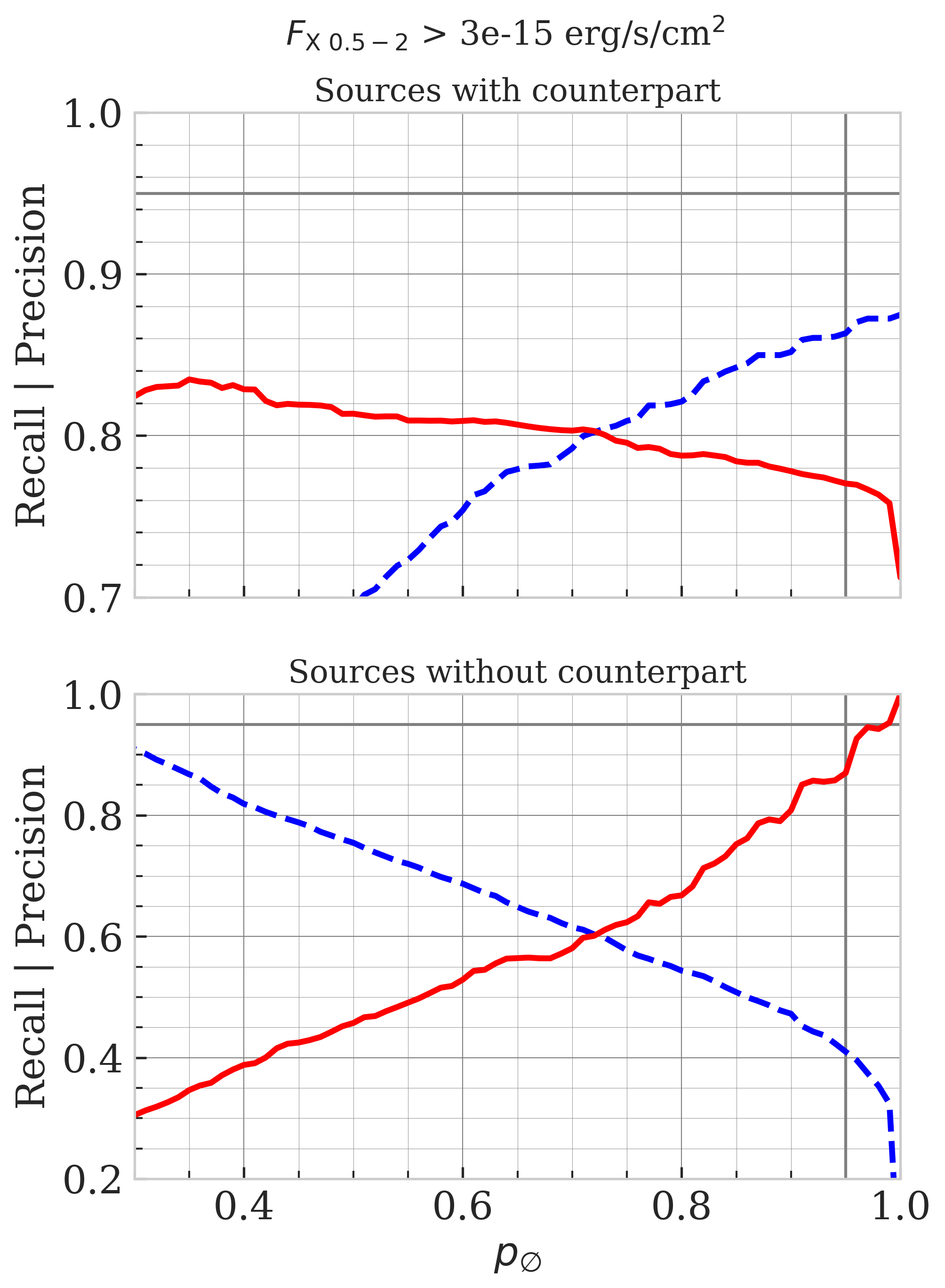}
    \end{subfigure}
    \caption{Selection quality metrics for the X-ray sources with an identified counterpart and hostless X-ray sources. The solid line shows precision for the sources with counterparts ($\text{Precision}_{\rm c}$, upper panels) and for the sources without counterparts ($\text{Precision}_{\rm h}$, lower panels). The dashed line corresponds to the selection recall for the sources with counterparts ($\text{Recall}_{\rm c}$, upper panels) and for the hostless sources ($\text{Recall}_{\rm h}$, lower panels). The metrics were calculated for the DESI LIS for two different X-ray flux thresholds.}
    \label{fig:metrics_pseudo}
\end{figure*}

\begin{itemize}
    \item Our cross-match algorithm allows searching optical counterparts for the LH \srge sources with precision 94\% and recall 94\% ($F_{\rm X, 0.5 - 2} > 10^{-14}$~\xdim\, the solid and dashed line on Fig.~\ref{fig:metrics_pseudo}, upper left panel). For an X-ray flux higher than $3\times 10^{-15}$ precision 77\% while recall 86\% (the solid and the dashed line, upper right panel). These values refer to $p_{\varnothing} = 0.95$ (shown by the vertical line on Fig.~\ref{fig:metrics_pseudo}) and for the DESI LIS.

    \item The algorithm allows selecting X-ray sources without optical counterparts (hostless sources) in the DESI LIS. For $F_{\rm X, 0.5 - 2} > 10^{-14}$~\xdim{} precision 77\% while recall 74\% (the solid and dashed line on Fig.~\ref{fig:metrics_pseudo}, lower left panel). For the flux higher than $3\times 10^{-15}$ precision 87\% while recall 41\% (the solid and dashed line on Fig.~\ref{fig:metrics_pseudo}, lower right panel). These values refer to $p_{\varnothing} = 0.95$.
    
\end{itemize}

\section{Conclusions} \label{conclusions}

We present a probabilistic cross-match model for the \srge sources detected in the Lockman Hole (LH) area. To illustrate the method we use the DESI Legacy Imaging Surveys optical data. To measure the effectiveness of our method we use the X-ray data from the CSC 2.0 and 4XMM DR10 catalogues.

The main results are listed bellow:

\begin{enumerate}

    \item Using the optical data from the DESI LIS in the fields of X-ray sources we measured the relationship between the localisation error of \srge LH sources and their detection likelihood $\sigma(\Lb)$. We found that the average localisation error of an X-ray source $\sigma_{\rm det}$ (calculated using the X-ray source detection algorithm eSASS) can be corrected with the following expresion: $\sigma_{\rm corr} = a\sqrt{ \sigma_{\rm det}^{b} + c^2}$, where $a = 0.87\pm0.08,\ b=2.53\pm0.20,\ c=1.12\pm0.19$. The error is measured in arcseconds.
    
    \item We measured the relationship $p_{\rm c}(F_{\rm X, 0.5 - 2})$ between the probability for an X-ray source (with a given flux) to have a counterpart in two photometric catalogues: the DESI LIS and SDSS. This probability equals $\approx74$\% and $\approx$45\% for X-ray sources with $F_{\rm X,0.5-2}\approx~\num{3e-15}$\,\xdim (for the DESI LIS and SDSS, respectively). For $F_{\rm X,0.5-2}\approx~10^{-14}$\,\xdim{} (corresponds to the expected four-year equatorial sensitivity for \srge) $p_{\rm c}$ values are similar both for the DESI LIS and SDSS: $\approx82$\% and $\approx75$\%. 
    
    \item Our model is able to take into account both positional and photometric information on optical candidates. We use photometric information in the form of effective magnitude. This approach allows reducing the number of counterparts missed during the selection process by $\approx$13\% (for X-ray sources with $F_{\rm X, 0.5 - 2} > \num{3e-15}$~\xdim{}) in comparison with the case when only positional information is used. The overall precision of the cross-match procedure for the \srge LH sources with $F_{\rm X, 0.5 - 2} > 3\times 10^{-15}$~\xdim{} is 78\%. For sources with flux higher than $10^{-14}$ --- 93\%. To calculate the overall precision, we define a correct match either as a correctly identified counterpart or as a correctly revealed X-ray source without a counterpart. Precision and recall vary depending on the value of $p_{\varnothing}$ (the probability that there is no counterpart for an X-ray source in the DESI LIS considering the vicinity of this X-ray source).
    
    \item The model allows searching optical counterparts for the LH \srge sources with precision 94\% and recall 94\% ($F_{\rm X, 0.5 - 2} > 10^{-14}$~\xdim{}). For an X-ray flux higher than $3\times 10^{-15}$ precision 77\% while recall 86\%. These values refer to $p_{\varnothing} = 0.95$ and for the DESI LIS.
    
    \item The algorithm allows selecting X-ray sources without optical counterparts (hostless sources) in the DESI LIS. For $F_{\rm X, 0.5 - 2} > 10^{-14}$~\xdim{} precision 77\% while recall 74\%. For the flux higher than $3\times 10^{-15}$ precision 87\% while recall 41\%. These values refer to $p_{\varnothing} = 0.95$.
\end{enumerate}

The model will be used in the whole-sky \srge survey to cross-match detected X-ray sources with different photometric catalogues.

\section*{}

This work is based on observations with the eROSITA telescope onboard the SRG observatory. The SRG observatory was built by Roskosmos in the interests of the Russian Academy of Sciences represented by its Space Research Institute (IKI) in the framework of the Russian Federal Space Program, with the participation of the Deutsches Zentrum für Luft- und Raumfahrt (DLR). The SRG/eROSITA X-ray telescope was built by a consortium of German Institutes led by MPE, and supported by DLR. The SRG spacecraft was designed, built, launched, and is operated by the Lavochkin Association and its subcontractors. The science data are downlinked via the Deep Space Network Antennae in Bear Lakes, Ussurijsk, and Baykonur, funded by Roskosmos. The eROSITA data used in this work were processed using the eSASS software system developed by the German eROSITA consortium and proprietary data reduction and analysis software developed by the Russian eROSITA Consortium. The SRGz system was developed in the Department of High Energy Astrophysics at IKI RAS.

This work benefited from the following publicly available software: Programming language Python, including
NumPy \citep{harris2020array} \& SciPy \citep{2020SciPy-NMeth}, Astropy \citep{astropy:2013, astropy:2018}, Pandas \citep{mckinney-proc-scipy-2010}, 
Matplotlib \citep{Hunter:2007}, CSCview\footnote{\url{http://cda.cfa.harvard.edu/cscview/}}, the TOPCAT analysis program \citep{2005ASPC..347...29T} and the SciServer\footnote{\url{www.sciserver.org}} scientific platform.

This work was supported by grant 21-12-00343 from the Russian Science Foundation.

\bibliographystyle{astl}
\bibliography{refs_rus.bib}


\newpage

\section*{Appendix}

\begin{table*}
    \renewcommand{\arraystretch}{2} 
    \begin{tabular}{lrrrrrrrrrrrrrrrr}
    \toprule
    № &    ${\Lb}_{\rm left}$ &  ${\Lb}_{\rm right}$ &  $\widetilde{\Lb}$ & $\tilde{\sigma}_{\rm det}$ & $\sigma$ & $\rho$ & $p_{\rm c}$ & $m_{\rm eff}$ & $\La$ &  $n_{\rm optical}$ &  $n_{\rm xray}$ \\
    \midrule
    0 &    6.00 &    10.44 &       7.88 &   3.32 &   $4.28_{-0.22}^{+0.22}$  &  $9.11_{-0.16}^{+0.06}$ &  $0.59_{-0.05}^{+0.03}$  &  23.15 & -59362.77 &  12796 &  1653 \\
    1 &   10.44 &    18.17 &      13.70 &   2.70 &   $3.22_{-0.14}^{+0.11}$  &  $7.84_{-0.09}^{+0.05}$ &  $0.63_{-0.04}^{+0.03}$  &  22.96 & -42495.91 &   9762 &  1438 \\
    2 &   18.17 &    31.62 &      23.63 &   2.20 &   $2.43_{-0.08}^{+0.06}$  &  $6.42_{-0.10}^{+0.05}$ &  $0.66_{-0.03}^{+0.02}$  &  22.68 & -27814.70 &   6950 &  1219 \\
    3 &   31.62 &    55.24 &      41.48 &   1.79 &   $2.18_{-0.09}^{+0.07}$  &  $5.55_{-0.14}^{+0.07}$ &  $0.70_{-0.04}^{+0.02}$  &  22.45 & -16972.17 &   4716 &   933 \\
    4 &   55.24 &    96.50 &      70.98 &   1.45 &   $1.76_{-0.07}^{+0.05}$  &  $2.74_{-0.11}^{+0.04}$ &  $0.72_{-0.04}^{+0.03}$  &  21.54 &  -5088.33 &   1868 &   651 \\
    5 &   96.50 &   168.58 &     118.30 &   1.17 &   $1.45_{-0.07}^{+0.05}$  &  $2.35_{-0.14}^{+0.07}$ &  $0.77_{-0.05}^{+0.04}$  &  21.32 &  -2245.76 &   1025 &   394 \\
    6 &  168.58 &   294.49 &     216.14 &   0.95 &   $1.23_{-0.07}^{+0.07}$  &  $2.03_{-0.18}^{+0.11}$ &  $0.77_{-0.06}^{+0.06}$  &  20.92 &   -864.96 &    544 &   231 \\
    7 &  294.49 &   514.45 &     374.95 &   0.78 &   $1.10_{-0.06}^{+0.06}$  &  $1.62_{-0.13}^{+0.09}$ &  $0.84_{-0.08}^{+0.08}$  &  20.45 &   -360.85 &    307 &   146 \\
    8 &  514.45 &   898.70 &     626.75 &   0.62 &   $1.14_{-0.10}^{+0.09}$  &  $0.81_{-0.16}^{+0.12}$ &  $0.88_{-0.12}^{+0.08}$  &  19.99 &    -23.92 &     90 &    60 \\
    9 &  898.70 &  1569.94 &    1075.39 &   0.50 &   $0.96_{-0.11}^{+0.12}$  &  $0.81_{-0.19}^{+0.17}$ &  $0.86_{-0.14}^{+0.08}$  &  19.59 &     4.97 &     62 &    39 \\
    \bottomrule
    \end{tabular}
\begin{center}
\caption{Limits for the detection likelihood ($\Lb$) bins and median values of $\Lb$ in these bins are shown in columns ${\Lb}_{\rm left}$, ${\Lb}_{\rm right}$ and $\widetilde{\Lb}$. Column $\tilde{\sigma}_{\rm det}$ contains the median values of the localisation error $\sigma_{\rm det}$. In columns $\sigma$,\  $\rho$,\  $p_{\rm c}$ model parameters are presented with their errors; $\La$ is for the maximum likelihood function values used in \S\ref{chap:pos_model}; $m_{\rm eff}$ is for the effective magnitude thresholds in corresponding bins; $n_{\rm xray}$ is a total number of the X-ray sources; $n_{\rm optical}$ is a total number of the optical sources taken form the fields of X-ray sources (within \ang{;;30}). All model parameters are calculated for $p_{\rm lim} = 0.85$ (see \S\ref{chap:plim_model}).}
\label{table:pars_table}
\end{center}
\end{table*}

\end{document}